\documentclass[zpreprint,zbstdefault]{zeus_paper}
\usepackage[english]{babel}

\newcommand{\Zpsrap}{%
The pseudorapidity is defined as $\eta=-\ln\left(\tan\frac{\theta}{2}\right)$,
where the polar angle, $\theta$, is measured with respect to the proton beam
direction.\xspace}

\newcommand{\Zdetdesc}{%
A detailed description of the ZEUS detector can be found 
elsewhere~\cite{zeus:1993:bluebook}. A brief outline of the 
components that are most relevant for this analysis is given
below.\xspace}

\newcommand{\Zsttdesc}[1]{%
The STT consisted of 48 sectors of two different sizes. Each sector
contained 192 (small sector) or 264 (large sector) straws of diameter
7.5 mm arranged into 3 layers. The sectors were trapezoidal in shape
and each subtended an azimuthal angle of $60^{\circ}$ -- 6 sectors
formed a so-called superlayer. A particle passing through the complete
detector traversed 8 superlayers, which were rotated around the beam
direction at angles of $30^{\circ}$ or $15^{\circ}$ to each other. The STT
covered the polar-angle region $5^{\circ}<\theta<23^{\circ}$.
}
\newcommand{\Zacknowledge}{%
We appreciate the contributions to the construction and maintenance of the 
ZEUS de- tector of many people who are not listed as authors. The HERA 
machine group and the DESY computing staff are especially acknowledged for 
their success in providing excel- lent operation of the collider and the 
data-analysis environment. We thank the DESY directorate for their strong 
support and encouragement.}

\chardef\usc=95
\chardef\til=126
\catcode`\@=11 
\DeclareRobustCommand\xdotspace{\futurelet\@let@token\@xdotspace}
\def\@xdotspace{%
  \ifx\@let@token.\else
  \ifx\@let@token\bgroup.\else
  \ifx\@let@token\egroup.\else
  \ifx\@let@token\/.\else
  \ifx\@let@token\ .\else
  \ifx\@let@token~.\else
  \ifx\@let@token!.\else
  \ifx\@let@token,.\else
  \ifx\@let@token:.\else
  \ifx\@let@token;.\else
  \ifx\@let@token?.\else
  \ifx\@let@token/.\else
  \ifx\@let@token'.\else
  \ifx\@let@token).\else
  \ifx\@let@token-.\else
  \ifx\@let@token\@xobeysp.\else
  \ifx\@let@token\space.\else
  \ifx\@let@token\@sptoken.\else
   .\space
   \fi\fi\fi\fi\fi\fi\fi\fi\fi\fi\fi\fi\fi\fi\fi\fi\fi\fi}
\catcode`\@=12 

\newcommand{\stru}[2]{%
   \relax\ifmmode\hbox{\vrule height#1 depth#2 width0pt}%
   \else\vrule height#1 depth#2 width0pt\fi}

\newcommand{\Ronum}[1]{\uppercase\expandafter{\romannumeral#1}}
\newcommand{\ronum}[1]{\expandafter{\romannumeral#1}}
\DeclareRobustCommand{\LaTeXZ}{%
  \LaTeX\kern-.05em4\kern-.1em
  {\raisebox{-0.2ex}{$\scriptstyle\text{ZEUS}$}}\xspace}

\newcommand{\eq}[1]{(\ref{eq-#1})}

\newcommand{\fig}[1]{Fig.~\ref{fig-#1}}
\newcommand{\Fig}[1]{Figure~\ref{fig-#1}}
\newcommand{\figand}[2]{Figs.~\ref{fig-#1} and~\ref{fig-#2}}

\newcommand{\tab}[1]{Table~\ref{tab-#1}}

\newcommand{\taband}[2]{Tables~\ref{tab-#1} and~\ref{tab-#2}}

\newcommand{\Sect}[1]{Section~\ref{sec-#1}}

\DeclareMathAlphabet{\mathbf}{OT1}{cmr}{bx}{sl}
\newcommand{\eVdist}{\kern-0.06667em}

\newcommand{\Gev}{{\text{Ge}\eVdist\text{V\/}}}

\newcommand{\mev}{{\,\text{Me}\eVdist\text{V\/}}}
\newcommand{\gev}{{\,\text{Ge}\eVdist\text{V\/}}}

\newcommand{\nb}{\,\text{nb}}

\newcommand{\pbi}{\,\text{pb}^{-1}}

\newcommand{\mum}{\,\upmu\text{m}}

\newcommand{\cm}{\,\text{cm}}

\newcommand{\Tesla}{\,\text{T}}

\newcommand{\slashfrac}[2]{%
  \raisebox{0.5ex}{\ensuremath #1}\kern-0.12em/\kern-0.08em
  \raisebox{-.8ex}{\ensuremath #2}}

\newcommand{\sqr}[3]{%
    {\vcenter{\hrule height.#3ex\hbox{\vrule width.#2ex height#1ex
     \kern#1ex\vrule width.#3ex}\hrule height.#2ex}}}

\catcode`\@=11 
\newcommand{\parenbar}{\mathpalette\p@renb@r}
\def\p@renb@r#1#2{\vbox{%
  \ifx#1\scriptscriptstyle \dimen@.7em\dimen@ii.2em\else
  \ifx#1\scriptstyle \dimen@.8em\dimen@ii.25em\else
  \dimen@1em\dimen@ii.4em\fi\fi \offinterlineskip
  \ialign{\hfill##\hfill\cr
    \vbox{\hrule width\dimen@ii}\cr
    \noalign{\vskip-.3ex}%
    \hbox to\dimen@{$\mathchar300\hfil\mathchar301$}\cr
    \noalign{\vskip-.3ex}%
    $#1#2$\cr}}}
\catcode`\@=12 

\newcommand{\als}{\alpha_s}

\newcommand{\rnge}{\hbox{$\,\text{--}\,$}}

\newcommand{\DA}{{\rm DA}}

\newcommand{\IP}{{\rm I$\kern-0.01667em$P}\xspace}

\newcommand{\Lumi}{{\cal L}}

\newcommand{\ord}[1]{{\cal O}(#1)}

\mathchardef\qsm=63
\mathchardef\pls=43
\mathchardef\mns=512
\mathchardef\plm=518
\mathchardef\eql=61
\mathchardef\smallleft=300
\mathchardef\smallright=301
\mathchardef\les=316
\mathchardef\gre=318
\mathchardef\leq=532
\mathchardef\grq=533

\catcode`\@=11 
\newcounter{pict@width}
\newcounter{pict@height}
\newlength{\pict@scale}
\newcommand{\psfigadd}[4]{%
\setcounter{pict@width}{1*\ratio{#2+\pict@scale/2}{\pict@scale}}
\setcounter{pict@height}{1*\ratio{#3+\pict@scale/2}{\pict@scale}}
\setlength{\unitlength}{\pict@scale}
\hbox to #2{\hspace{-\fill}\begin{picture}(\thepict@width,\thepict@height)
\put(0,0){\psfig{figure=#1,width=#2,height=#3,clip=}}
\SetScale{0.283466457}
\SetWidth{1.763889}
{#4}
\end{picture}}
}
\newcounter{pict@widthfst}
\newcounter{pict@widthscd}
\newcounter{pict@widthtot}
\newcommand{\psfigaddtwo}[7]{%
\setcounter{pict@widthfst}{1*\ratio{#2+\pict@scale/2}{\pict@scale}}
\setcounter{pict@widthscd}{1*\ratio{#2+#4+\pict@scale/2}{\pict@scale}}
\setcounter{pict@widthtot}{1*\ratio{#2+#4+#6+\pict@scale/2}{\pict@scale}}
\setcounter{pict@height}{1*\ratio{#3+\pict@scale/2}{\pict@scale}}
\setlength{\unitlength}{\pict@scale}
\hbox{\hspace{-\fill}\begin{picture}(\thepict@widthtot,\thepict@height)
\put(0,0){\psfig{figure=#1,width=#2,height=#3,clip=}}
\put(\thepict@widthscd,0){\psfig{figure=#5,width=#6,height=#3,clip=}}
\SetScale{0.283466457}
\SetWidth{1.763889}
{#7}
\end{picture}}
}
\newcommand{\psfigror}[4]{%
\setcounter{pict@width}{1*\ratio{#2+\pict@scale/2}{\pict@scale}}
\setcounter{pict@height}{1*\ratio{#3+\pict@scale/2}{\pict@scale}}
\setlength{\unitlength}{\pict@scale}
\hbox{\begin{picture}(\thepict@width,\thepict@height)
\put(0,\thepict@height){\psfig{figure=#1,width=#3,height=#2,clip=,angle=270}}
\SetScale{0.283466457}
\SetWidth{1.763889}
{#4}
\end{picture}}
}
\newcommand{\psfigrol}[4]{%
\setcounter{pict@width}{1*\ratio{#2+\pict@scale/2}{\pict@scale}}
\setcounter{pict@height}{1*\ratio{#3+\pict@scale/2}{\pict@scale}}
\setlength{\unitlength}{\pict@scale}
\hbox{\begin{picture}(\thepict@width,\thepict@height)
\put(0,0){\psfig{figure=#1,width=#3,height=#2,clip=,angle=90}}
\SetScale{0.283466457}
\SetWidth{1.763889}
{#4}
\end{picture}}
}
\catcode`\@=12 
\newlength\listtextwidth

\newcommand{\pcite}[1]{{\protect\cite{#1}}}

\catcode`\@=11 
\newlength{\@tabfninsert}
\newlength{\@tabfnwidth}
\newcommand{\tabfootnote}[2]{%
  \setlength{\@tabfninsert}{0.8em}
  \setlength{\@tabfnwidth}{\textwidth}
  \addtolength{\@tabfnwidth}{-\@tabfninsert}
  \addtolength{\@tabfnwidth}{-0.4em}
  \noindent\makebox[\@tabfninsert][r]{\footnotesize$^{#1}$\hfil}\hfill%
  \parbox[t]{\@tabfnwidth}{\footnotesize #2\hfill}}
\catcode`\@=12 

\newcommand{\hvqdis}{\textsc{HVQDIS}\xspace}
\newcommand{\HVQDIS}{\textsc{HVQDIS}\xspace}

\newcommand{\HERACLES}{\textsc{HERACLES}\xspace}

\newcommand{\RAPGAP}{\textsc{RAPGAP}\xspace}

\newcommand{\rmrad}{\mathrm{rad}}
\newcommand{\rmBorn}{\mathrm{Born}}

\newcommand{\ftwocc}{$F_2^{c\bar{c}}$\xspace}

\newcommand{\alsmzone}{{\alpha_s(M_Z)}}

\newcommand{\Zctdmvddescnew}[1]{%
In the kinematic range of the analysis, charged particles were tracked
in the central tracking detector (CTD)~\citeCTD and the microvertex
detector (MVD)~\citeMVD. These components operated in a magnetic
field of $1.43\Tesla$ provided by a thin superconducting solenoid. The
CTD consisted of 72~cylindrical drift-chamber layers, organised in nine
superlayers covering the polar-angle#1 region
\mbox{$15^\circ<\theta<164^\circ$}.
The MVD silicon tracker consisted of a barrel (BMVD) and a forward
(FMVD) section. The BMVD contained three layers and provided
polar-angle coverage for tracks from $30^\circ$ to
$150^\circ$. The four-layer FMVD extended the polar-angle coverage in
the forward region to $7^\circ$. After alignment, the single-hit
resolution of the MVD was $\rm 24\mum$. The transverse distance of closest
approach (DCA) of tracks to the nominal vertex in $XY$ was measured to have
a resolution, averaged over the azimuthal angle, of $(46 \oplus 122 /
p_{T})\mum$, with $p_{T}$ in $\Gev$.  For CTD-MVD tracks that pass
through all nine CTD superlayers, the momentum resolution was
$\sigma(p_{T})/p_{T} = 0.0029 p_{T} \oplus 0.0081 \oplus
0.0012/p_{T}$, with $p_{T}$ in $\Gev$.}
\newcommand{\Zcaldescnew}{%
The high-resolution uranium--scintillator calorimeter (CAL)~\citeCAL consisted 
of three parts: the forward (FCAL), the barrel (BCAL) and the rear (RCAL)
calorimeters. Each part was subdivided transversely into towers and
longitudinally into one electromagnetic section (EMC) and either one (in RCAL)
or two (in BCAL and FCAL) hadronic sections (HAC). The smallest subdivision of
the calorimeter is called a cell.  The CAL energy resolutions, as measured under
test-beam conditions, are $\sigma(E)/E=0.18/\sqrt{E}$ for electrons and
$\sigma(E)/E=0.35/\sqrt{E}$ for hadrons, with $E$ in $\Gev$.}
\newcommand{\Zlumidescnew}[1]{%
The luminosity was measured using the Bethe-Heitler reaction
$ep\,\rightarrow\, e\gamma p$ by a luminosity detector which consisted
of independent lead--scintillator calorimeter\citePCAL and magnetic
spectrometer\citeSPECTRO systems. The fractional systematic
uncertainty on the measured luminosity was #1.}
\newcommand{\ZcoosysBnew}{%
The ZEUS coordinate system is a right-handed Cartesian system, with the $Z$
axis pointing in the proton beam direction, referred to as the ``forward
direction'', and the $X$ axis pointing towards the centre of HERA.
The coordinate origin is at the nominal interaction point.\xspace}
\newcommand{\ZcoosysfnBetanew}{\footnote{\ZcoosysBnew\Zpsrap}}

\def\citeCTD{{\cite{%
nim:a279:290,*npps:b32:181,*nim:a338:254%
}}\xspace}
\def\citeMVD{{\cite{%
nim:a581:656%
}}\xspace}

\def\citeCAL{{\cite{%
nim:a309:77,*nim:a309:101,*nim:a321:356,*nim:a336:23%
}}\xspace}

\def\cite6mT{{\cite{%
thesis:gosau:2007%
}}\xspace}

\def\citePCAL{{\cite{%
desy-92-066,*zfp:c63:391,*acpp:b32:2025%
}}\xspace}
\def\citeSPECTRO{{\cite{%
nim:a565:572%
}}\xspace}
\def\citeBeauty{{\cite{%
epj:c65:65-79,epj:c69:347,epj:c71:1573%
}}\xspace}

\begin{document}
\hyphenation{RAPGAP HERACLES}
\graphicspath{{./figs/}}
\prepnum{DESY--13--028}

\title{Measurement of \boldmath{$D^{\pm}$} production in deep inelastic $ep$ scattering with the \textsc{ZEUS} detector at \textsc{HERA}}                                                       
                    
\author{ZEUS Collaboration}

\abstract{
Charm production in deep inelastic \emph{ep} scattering was measured with the ZEUS detector using 
an integrated luminosity of $354\pbi$.
Charm quarks were identified by reconstructing $D^{\pm}$ mesons in the $D^{\pm} \rightarrow K^{\mp}\pi^{\pm}\pi^{\pm}$ decay channel. Lifetime information was used to reduce combinatorial background substantially. Differential cross sections were measured in the kinematic region  \mbox{$5 < Q^{2} < 1000\gev^{2}$}, \mbox{$0.02 < y < 0.7$}, \mbox{$1.5 < p_{T}(D^{\pm}) < 15\gev$} and \mbox{$|\eta(D^{\pm})| < 1.6$}, where $Q^2$ is the photon virtuality, $y$ is the inelasticity, and $p_T(D^{\pm})$ and $\eta(D^{\pm})$ are the transverse momentum and the pseudorapidity of the $D^{\pm}$ meson, respectively. Next-to-leading-order QCD predictions are compared to the data.  The charm contribution, \ftwocc, to the proton structure-function $F_2$ was extracted.}
\makezeustitle

\pagestyle{plain}
\clearpage

%
%
%
%

\pagenumbering{Roman}
                                                   %
\begin{center}
{                      \Large  The ZEUS Collaboration              }
\end{center}

{\small


        {\raggedright
H.~Abramowicz$^{45, aj}$, 
I.~Abt$^{35}$, 
L.~Adamczyk$^{13}$, 
M.~Adamus$^{54}$, 
R.~Aggarwal$^{7, c}$, 
S.~Antonelli$^{4}$, 
P.~Antonioli$^{3}$, 
A.~Antonov$^{33}$, 
M.~Arneodo$^{50}$, 
O.~Arslan$^{5}$
% ...
        }

\newpage


\makebox[3em]{$^{1}$}
\begin{minipage}[t]{14cm}
{\it Argonne National Laboratory, Argonne, Illinois 60439-4815, USA}~$^{A}$

\end{minipage}\\
\makebox[3em]{$^{2}$}
\begin{minipage}[t]{14cm}
{\it Andrews University, Berrien Springs, Michigan 49104-0380, USA}

\end{minipage}\\
\makebox[3em]{$^{3}$}
\begin{minipage}[t]{14cm}
{\it INFN Bologna, Bologna, Italy}~$^{B}$

\end{minipage}\\
\makebox[3em]{$^{4}$}
\begin{minipage}[t]{14cm}
{\it University and INFN Bologna, Bologna, Italy}~$^{B}$

\end{minipage}\\
\makebox[3em]{$^{5}$}
\begin{minipage}[t]{14cm}
{\it Physikalisches Institut der Universit\"at Bonn,
Bonn, Germany}~$^{C}$

\end{minipage}\\
\makebox[3em]{$^{6}$}
\begin{minipage}[t]{14cm}
{\it H.H.~Wills Physics Laboratory, University of Bristol,
Bristol, United Kingdom}~$^{D}$

\end{minipage}\\
\makebox[3em]{$^{7}$}
\begin{minipage}[t]{14cm}
{\it Panjab University, Department of Physics, Chandigarh, India}

\end{minipage}\\
\makebox[3em]{$^{8}$}
\begin{minipage}[t]{14cm}
{\it Calabria University,
Physics Department and INFN, Cosenza, Italy}~$^{B}$

\end{minipage}\\
\makebox[3em]{$^{9}$}
\begin{minipage}[t]{14cm}
{\it Institute for Universe and Elementary Particles, Chonnam National University,\\
Kwangju, South Korea}

\end{minipage}\\
\makebox[3em]{$^{10}$}
\begin{minipage}[t]{14cm}
{\it Jabatan Fizik, Universiti Malaya, 50603 Kuala Lumpur, Malaysia}~$^{E}$

\end{minipage}\\
\makebox[3em]{$^{11}$}
\begin{minipage}[t]{14cm}
{\it Nevis Laboratories, Columbia University, Irvington on Hudson,
New York 10027, USA}~$^{F}$

\end{minipage}\\
\makebox[3em]{$^{12}$}
\begin{minipage}[t]{14cm}
{\it The Henryk Niewodniczanski Institute of Nuclear Physics, Polish Academy of \\
Sciences, Krakow, Poland}~$^{G}$

\end{minipage}\\
\makebox[3em]{$^{13}$}
\begin{minipage}[t]{14cm}
{\it AGH-University of Science and Technology, Faculty of Physics and Applied Computer
Science, Krakow, Poland}~$^{H}$

\end{minipage}\\
\makebox[3em]{$^{14}$}
\begin{minipage}[t]{14cm}
{\it Department of Physics, Jagellonian University, Cracow, Poland}

\end{minipage}\\
\makebox[3em]{$^{15}$}
\begin{minipage}[t]{14cm}
{\it Deutsches Elektronen-Synchrotron DESY, Hamburg, Germany}

\end{minipage}\\
\makebox[3em]{$^{16}$}
\begin{minipage}[t]{14cm}
{\it Deutsches Elektronen-Synchrotron DESY, Zeuthen, Germany}

\end{minipage}\\
\makebox[3em]{$^{17}$}
\begin{minipage}[t]{14cm}
{\it INFN Florence, Florence, Italy}~$^{B}$

\end{minipage}\\
\makebox[3em]{$^{18}$}
\begin{minipage}[t]{14cm}
{\it University and INFN Florence, Florence, Italy}~$^{B}$

\end{minipage}\\
\makebox[3em]{$^{19}$}
\begin{minipage}[t]{14cm}
{\it Fakult\"at f\"ur Physik der Universit\"at Freiburg i.Br.,
Freiburg i.Br., Germany}

\end{minipage}\\
\makebox[3em]{$^{20}$}
\begin{minipage}[t]{14cm}
{\it School of Physics and Astronomy, University of Glasgow,
Glasgow, United Kingdom}~$^{D}$

\end{minipage}\\
\makebox[3em]{$^{21}$}
\begin{minipage}[t]{14cm}
{\it Department of Engineering in Management and Finance, Univ. of
the Aegean, Chios, Greece}

\end{minipage}\\
\makebox[3em]{$^{22}$}
\begin{minipage}[t]{14cm}
{\it Hamburg University, Institute of Experimental Physics, Hamburg,
Germany}~$^{I}$

\end{minipage}\\
\makebox[3em]{$^{23}$}
\begin{minipage}[t]{14cm}
{\it Imperial College London, High Energy Nuclear Physics Group,
London, United Kingdom}~$^{D}$

\end{minipage}\\
\makebox[3em]{$^{24}$}
\begin{minipage}[t]{14cm}
{\it Institute of Particle and Nuclear Studies, KEK,
Tsukuba, Japan}~$^{J}$

\end{minipage}\\
\makebox[3em]{$^{25}$}
\begin{minipage}[t]{14cm}
{\it Institute of Physics and Technology of Ministry of Education and
Science of Kazakhstan, Almaty, Kazakhstan}

\end{minipage}\\
\makebox[3em]{$^{26}$}
\begin{minipage}[t]{14cm}
{\it Institute for Nuclear Research, National Academy of Sciences, Kyiv, Ukraine}

\end{minipage}\\
\makebox[3em]{$^{27}$}
\begin{minipage}[t]{14cm}
{\it Department of Nuclear Physics, National Taras Shevchenko University of Kyiv, Kyiv, Ukraine}

\end{minipage}\\
\makebox[3em]{$^{28}$}
\begin{minipage}[t]{14cm}
{\it Kyungpook National University, Center for High Energy Physics, Daegu,
South Korea}~$^{K}$

\end{minipage}\\
\makebox[3em]{$^{29}$}
\begin{minipage}[t]{14cm}
{\it Institut de Physique Nucl\'{e}aire, Universit\'{e} Catholique de Louvain, Louvain-la-Neuve,\\
Belgium}~$^{L}$

\end{minipage}\\
\makebox[3em]{$^{30}$}
\begin{minipage}[t]{14cm}
{\it Departamento de F\'{\i}sica Te\'orica, Universidad Aut\'onoma
de Madrid, Madrid, Spain}~$^{M}$

\end{minipage}\\
\makebox[3em]{$^{31}$}
\begin{minipage}[t]{14cm}
{\it Department of Physics, McGill University,
Montr\'eal, Qu\'ebec, Canada H3A 2T8}~$^{N}$

\end{minipage}\\
\makebox[3em]{$^{32}$}
\begin{minipage}[t]{14cm}
{\it Meiji Gakuin University, Faculty of General Education,
Yokohama, Japan}~$^{J}$

\end{minipage}\\
\makebox[3em]{$^{33}$}
\begin{minipage}[t]{14cm}
{\it Moscow Engineering Physics Institute, Moscow, Russia}~$^{O}$

\end{minipage}\\
\makebox[3em]{$^{34}$}
\begin{minipage}[t]{14cm}
{\it Lomonosov Moscow State University, Skobeltsyn Institute of Nuclear Physics,
Moscow, Russia}~$^{P}$

\end{minipage}\\
\makebox[3em]{$^{35}$}
\begin{minipage}[t]{14cm}
{\it Max-Planck-Institut f\"ur Physik, M\"unchen, Germany}

\end{minipage}\\
\makebox[3em]{$^{36}$}
\begin{minipage}[t]{14cm}
{\it NIKHEF and University of Amsterdam, Amsterdam, Netherlands}~$^{Q}$

\end{minipage}\\
\makebox[3em]{$^{37}$}
\begin{minipage}[t]{14cm}
{\it Physics Department, Ohio State University,
Columbus, Ohio 43210, USA}~$^{A}$

\end{minipage}\\
\makebox[3em]{$^{38}$}
\begin{minipage}[t]{14cm}
{\it Department of Physics, University of Oxford,
Oxford, United Kingdom}~$^{D}$

\end{minipage}\\
\makebox[3em]{$^{39}$}
\begin{minipage}[t]{14cm}
{\it INFN Padova, Padova, Italy}~$^{B}$

\end{minipage}\\
\makebox[3em]{$^{40}$}
\begin{minipage}[t]{14cm}
{\it Dipartimento di Fisica dell' Universit\`a and INFN,
Padova, Italy}~$^{B}$

\end{minipage}\\
\makebox[3em]{$^{41}$}
\begin{minipage}[t]{14cm}
{\it Department of Physics, Pennsylvania State University, University Park,\\
Pennsylvania 16802, USA}~$^{F}$

\end{minipage}\\
\makebox[3em]{$^{42}$}
\begin{minipage}[t]{14cm}
{\it Polytechnic University, Tokyo, Japan}~$^{J}$

\end{minipage}\\
\makebox[3em]{$^{43}$}
\begin{minipage}[t]{14cm}
{\it Dipartimento di Fisica, Universit\`a 'La Sapienza' and INFN,
Rome, Italy}~$^{B}$

\end{minipage}\\
\makebox[3em]{$^{44}$}
\begin{minipage}[t]{14cm}
{\it Rutherford Appleton Laboratory, Chilton, Didcot, Oxon,
United Kingdom}~$^{D}$

\end{minipage}\\
\makebox[3em]{$^{45}$}
\begin{minipage}[t]{14cm}
{\it Raymond and Beverly Sackler Faculty of Exact Sciences, School of Physics, \\
Tel Aviv University, Tel Aviv, Israel}~$^{R}$

\end{minipage}\\
\makebox[3em]{$^{46}$}
\begin{minipage}[t]{14cm}
{\it Department of Physics, Tokyo Institute of Technology,
Tokyo, Japan}~$^{J}$

\end{minipage}\\
\makebox[3em]{$^{47}$}
\begin{minipage}[t]{14cm}
{\it Department of Physics, University of Tokyo,
Tokyo, Japan}~$^{J}$

\end{minipage}\\
\makebox[3em]{$^{48}$}
\begin{minipage}[t]{14cm}
{\it Tokyo Metropolitan University, Department of Physics,
Tokyo, Japan}~$^{J}$

\end{minipage}\\
\makebox[3em]{$^{49}$}
\begin{minipage}[t]{14cm}
{\it Universit\`a di Torino and INFN, Torino, Italy}~$^{B}$

\end{minipage}\\
\makebox[3em]{$^{50}$}
\begin{minipage}[t]{14cm}
{\it Universit\`a del Piemonte Orientale, Novara, and INFN, Torino,
Italy}~$^{B}$

\end{minipage}\\
\makebox[3em]{$^{51}$}
\begin{minipage}[t]{14cm}
{\it Department of Physics, University of Toronto, Toronto, Ontario,
Canada M5S 1A7}~$^{N}$

\end{minipage}\\
\makebox[3em]{$^{52}$}
\begin{minipage}[t]{14cm}
{\it Physics and Astronomy Department, University College London,
London, United Kingdom}~$^{D}$

\end{minipage}\\
\makebox[3em]{$^{53}$}
\begin{minipage}[t]{14cm}
{\it Faculty of Physics, University of Warsaw, Warsaw, Poland}

\end{minipage}\\
\makebox[3em]{$^{54}$}
\begin{minipage}[t]{14cm}
{\it National Centre for Nuclear Research, Warsaw, Poland}

\end{minipage}\\
\makebox[3em]{$^{55}$}
\begin{minipage}[t]{14cm}
{\it Department of Particle Physics and Astrophysics, Weizmann
Institute, Rehovot, Israel}

\end{minipage}\\
\makebox[3em]{$^{56}$}
\begin{minipage}[t]{14cm}
{\it Department of Physics, University of Wisconsin, Madison,
Wisconsin 53706, USA}~$^{A}$

\end{minipage}\\
\makebox[3em]{$^{57}$}
\begin{minipage}[t]{14cm}
{\it Department of Physics, York University, Ontario, Canada M3J 1P3}~$^{N}$

\end{minipage}\\
\vspace{30em} \pagebreak[4]


\makebox[3ex]{$^{ A}$}
\begin{minipage}[t]{14cm}
 supported by the US Department of Energy\
\end{minipage}\\
\makebox[3ex]{$^{ B}$}
\begin{minipage}[t]{14cm}
 supported by the Italian National Institute for Nuclear Physics (INFN) \
\end{minipage}\\
\makebox[3ex]{$^{ C}$}
\begin{minipage}[t]{14cm}
 supported by the German Federal Ministry for Education and Research (BMBF), under
 contract No. 05 H09PDF\
\end{minipage}\\
\makebox[3ex]{$^{ D}$}
\begin{minipage}[t]{14cm}
 supported by the Science and Technology Facilities Council, UK\
\end{minipage}\\
\makebox[3ex]{$^{ E}$}
\begin{minipage}[t]{14cm}
 supported by HIR and UMRG grants from Universiti Malaya, and an ERGS grant from the
 Malaysian Ministry for Higher Education\
\end{minipage}\\
\makebox[3ex]{$^{ F}$}
\begin{minipage}[t]{14cm}
 supported by the US National Science Foundation. Any opinion, findings and conclusions or
 recommendations expressed in this material are those of the authors and do not necessarily
 reflect the views of the National Science Foundation.\
\end{minipage}\\
\makebox[3ex]{$^{ G}$}
\begin{minipage}[t]{14cm}
 supported by the Polish Ministry of Science and Higher Education as a scientific project No.
 DPN/N188/DESY/2009\
\end{minipage}\\
\makebox[3ex]{$^{ H}$}
\begin{minipage}[t]{14cm}
 supported by the Polish Ministry of Science and Higher Education and its grants
 for Scientific Research\
\end{minipage}\\
\makebox[3ex]{$^{ I}$}
\begin{minipage}[t]{14cm}
 supported by the German Federal Ministry for Education and Research (BMBF), under
 contract No. 05h09GUF, and the SFB 676 of the Deutsche Forschungsgemeinschaft (DFG) \
\end{minipage}\\
\makebox[3ex]{$^{ J}$}
\begin{minipage}[t]{14cm}
 supported by the Japanese Ministry of Education, Culture, Sports, Science and Technology
 (MEXT) and its grants for Scientific Research\
\end{minipage}\\
\makebox[3ex]{$^{ K}$}
\begin{minipage}[t]{14cm}
 supported by the Korean Ministry of Education and Korea Science and Engineering Foundation\
\end{minipage}\\
\makebox[3ex]{$^{ L}$}
\begin{minipage}[t]{14cm}
 supported by FNRS and its associated funds (IISN and FRIA) and by an Inter-University
 Attraction Poles Programme subsidised by the Belgian Federal Science Policy Office\
\end{minipage}\\
\makebox[3ex]{$^{ M}$}
\begin{minipage}[t]{14cm}
 supported by the Spanish Ministry of Education and Science through funds provided by CICYT\
\end{minipage}\\
\makebox[3ex]{$^{ N}$}
\begin{minipage}[t]{14cm}
 supported by the Natural Sciences and Engineering Research Council of Canada (NSERC) \
\end{minipage}\\
\makebox[3ex]{$^{ O}$}
\begin{minipage}[t]{14cm}
 partially supported by the German Federal Ministry for Education and Research (BMBF)\
\end{minipage}\\
\makebox[3ex]{$^{ P}$}
\begin{minipage}[t]{14cm}
 supported by RF Presidential grant N 3920.2012.2 for the Leading Scientific Schools and by
 the Russian Ministry of Education and Science through its grant for Scientific Research on
 High Energy Physics\
\end{minipage}\\
\makebox[3ex]{$^{ Q}$}
\begin{minipage}[t]{14cm}
 supported by the Netherlands Foundation for Research on Matter (FOM)\
\end{minipage}\\
\makebox[3ex]{$^{ R}$}
\begin{minipage}[t]{14cm}
 supported by the Israel Science Foundation\
\end{minipage}\\
\vspace{30em} \pagebreak[4]


\makebox[3ex]{$^{ a}$}
\begin{minipage}[t]{14cm}
now at University of Salerno, Italy\
\end{minipage}\\
\makebox[3ex]{$^{ b}$}
\begin{minipage}[t]{14cm}
now at Queen Mary University of London, United Kingdom\
\end{minipage}\\
\makebox[3ex]{$^{ c}$}
\begin{minipage}[t]{14cm}
also funded by Max Planck Institute for Physics, Munich, Germany\
\end{minipage}\\
\makebox[3ex]{$^{ d}$}
\begin{minipage}[t]{14cm}
also Senior Alexander von Humboldt Research Fellow at Hamburg University,
 Institute of Experimental Physics, Hamburg, Germany\
\end{minipage}\\
\makebox[3ex]{$^{ e}$}
\begin{minipage}[t]{14cm}
also at Cracow University of Technology, Faculty of Physics,
 Mathematics and Applied Computer Science, Poland\
\end{minipage}\\
\makebox[3ex]{$^{ f}$}
\begin{minipage}[t]{14cm}
supported by the research grant No. 1 P03B 04529 (2005-2008)\
\end{minipage}\\
\makebox[3ex]{$^{ g}$}
\begin{minipage}[t]{14cm}
partially supported by the Polish National Science Centre projects DEC-2011/01/B/ST2/03643
 and DEC-2011/03/B/ST2/00220\
\end{minipage}\\
\makebox[3ex]{$^{ h}$}
\begin{minipage}[t]{14cm}
now at Rockefeller University, New York, NY
 10065, USA\
\end{minipage}\\
\makebox[3ex]{$^{ i}$}
\begin{minipage}[t]{14cm}
now at DESY group FS-CFEL-1\
\end{minipage}\\
\makebox[3ex]{$^{ j}$}
\begin{minipage}[t]{14cm}
now at Institute of High Energy Physics, Beijing, China\
\end{minipage}\\
\makebox[3ex]{$^{ k}$}
\begin{minipage}[t]{14cm}
now at DESY group FEB, Hamburg, Germany\
\end{minipage}\\
\makebox[3ex]{$^{ l}$}
\begin{minipage}[t]{14cm}
also at Moscow State University, Russia\
\end{minipage}\\
\makebox[3ex]{$^{ m}$}
\begin{minipage}[t]{14cm}
now at University of Liverpool, United Kingdom\
\end{minipage}\\
\makebox[3ex]{$^{ n}$}
\begin{minipage}[t]{14cm}
now at CERN, Geneva, Switzerland\
\end{minipage}\\
\makebox[3ex]{$^{ o}$}
\begin{minipage}[t]{14cm}
also affiliated with University College London, UK\
\end{minipage}\\
\makebox[3ex]{$^{ p}$}
\begin{minipage}[t]{14cm}
now at Goldman Sachs, London, UK\
\end{minipage}\\
\makebox[3ex]{$^{ q}$}
\begin{minipage}[t]{14cm}
also at Institute of Theoretical and Experimental Physics, Moscow, Russia\
\end{minipage}\\
\makebox[3ex]{$^{ r}$}
\begin{minipage}[t]{14cm}
also at FPACS, AGH-UST, Cracow, Poland\
\end{minipage}\\
\makebox[3ex]{$^{ s}$}
\begin{minipage}[t]{14cm}
partially supported by Warsaw University, Poland\
\end{minipage}\\
\makebox[3ex]{$^{ t}$}
\begin{minipage}[t]{14cm}
supported by the Alexander von Humboldt Foundation\
\end{minipage}\\
\makebox[3ex]{$^{ u}$}
\begin{minipage}[t]{14cm}
now at Istituto Nazionale di Fisica Nucleare (INFN), Pisa, Italy\
\end{minipage}\\
\makebox[3ex]{$^{ v}$}
\begin{minipage}[t]{14cm}
now at Haase Energie Technik AG, Neum\"unster, Germany\
\end{minipage}\\
\makebox[3ex]{$^{ w}$}
\begin{minipage}[t]{14cm}
now at Department of Physics, University of Bonn, Germany\
\end{minipage}\\
\makebox[3ex]{$^{ x}$}
\begin{minipage}[t]{14cm}
also affiliated with DESY, Germany\
\end{minipage}\\
\makebox[3ex]{$^{ y}$}
\begin{minipage}[t]{14cm}
also at University of Tokyo, Japan\
\end{minipage}\\
\makebox[3ex]{$^{ z}$}
\begin{minipage}[t]{14cm}
now at Kobe University, Japan\
\end{minipage}\\
\makebox[3ex]{$^{\dagger}$}
\begin{minipage}[t]{14cm}
 deceased \
\end{minipage}\\
\makebox[3ex]{$^{aa}$}
\begin{minipage}[t]{14cm}
supported by DESY, Germany\
\end{minipage}\\
\makebox[3ex]{$^{ab}$}
\begin{minipage}[t]{14cm}
member of National Technical University of Ukraine, Kyiv Polytechnic Institute,
 Kyiv, Ukraine\
\end{minipage}\\
\makebox[3ex]{$^{ac}$}
\begin{minipage}[t]{14cm}
member of National Technical University of Ukraine, Kyiv, Ukraine\
\end{minipage}\\
\makebox[3ex]{$^{ad}$}
\begin{minipage}[t]{14cm}
now at DESY ATLAS group\
\end{minipage}\\
\makebox[3ex]{$^{ae}$}
\begin{minipage}[t]{14cm}
member of National University of Kyiv - Mohyla Academy, Kyiv, Ukraine\
\end{minipage}\\
\makebox[3ex]{$^{af}$}
\begin{minipage}[t]{14cm}
Alexander von Humboldt Professor; also at DESY and University of Oxford\
\end{minipage}\\
\makebox[3ex]{$^{ag}$}
\begin{minipage}[t]{14cm}
STFC Advanced Fellow\
\end{minipage}\\
\makebox[3ex]{$^{ah}$}
\begin{minipage}[t]{14cm}
now at LNF, Frascati, Italy\
\end{minipage}\\
\makebox[3ex]{$^{ai}$}
\begin{minipage}[t]{14cm}
This material was based on work supported by the
 National Science Foundation, while working at the Foundation.\
\end{minipage}\\
\makebox[3ex]{$^{aj}$}
\begin{minipage}[t]{14cm}
also at Max Planck Institute for Physics, Munich, Germany, External Scientific Member\
\end{minipage}\\
\makebox[3ex]{$^{ak}$}
\begin{minipage}[t]{14cm}
now at Tokyo Metropolitan University, Japan\
\end{minipage}\\
\makebox[3ex]{$^{al}$}
\begin{minipage}[t]{14cm}
now at Nihon Institute of Medical Science, Japan\
\end{minipage}\\
\makebox[3ex]{$^{am}$}
\begin{minipage}[t]{14cm}
now at Osaka University, Osaka, Japan\
\end{minipage}\\
\makebox[3ex]{$^{an}$}
\begin{minipage}[t]{14cm}
also at \L\'{o}d\'{z} University, Poland\
\end{minipage}\\
\makebox[3ex]{$^{ao}$}
\begin{minipage}[t]{14cm}
member of \L\'{o}d\'{z} University, Poland\
\end{minipage}\\
\makebox[3ex]{$^{ap}$}
\begin{minipage}[t]{14cm}
now at Department of Physics, Stockholm University, Stockholm, Sweden\
\end{minipage}\\
\makebox[3ex]{$^{aq}$}
\begin{minipage}[t]{14cm}
also at Cardinal Stefan Wyszy\'nski University, Warsaw, Poland\
\end{minipage}\\

}


\clearpage
\pagenumbering{arabic}
\section{Introduction}
\label{sec-int}

Measurements of charm production in deep inelastic \emph{ep} scattering (DIS) at HERA 
provide powerful constraints on the proton structure.
The dominant production mechanism, the boson-gluon fusion (BGF) process, $\gamma g \to c \bar{c}$, 
provides direct access to the gluon content of the proton.
As charm-quark production contributes up to $30\%$ of the inclusive DIS cross sections at HERA, 
a correct modelling of this contribution in perturbative QCD (pQCD) calculations is
important for determining parton distribution functions (PDFs) of the proton.
One crucial issue is the treatment of charm-quark mass effects.

At HERA, several different charm-tagging techniques have been exploited to measure 
charm production in DIS \cite{epj:c12:35, pr:d69:012004, epj:c63:2009:2:171-188, epj:c65:65-79, jhep:2010:11:009, pl:b528:199, epj:c51:271, pl:b686:91, epj:c71:1769, epj:c45:23, epj:c65:89, epj:c71:1509}.  
Recently, a combined analysis of these data was performed~\cite{tech:desy12-XXX:ccombination}, 
yielding results with both statistical and systematic uncertainties significantly reduced. 
In general, perturbative QCD predictions at next-to-leading order (NLO)  
are in reasonable agreement with the measurements.
These data have also been used to obtain a precise determination of the charm-quark mass~\cite{pl:b699:345, tech:desy12-XXX:ccombination}.

In the analysis presented here, a charm quark in the final state was identified by the presence of a $D^{+}$ meson\footnote{Charge-conjugate modes are implied throughout the paper.}, using the $D^{+} \rightarrow K^{-}\pi^{+}\pi^{+}$ decay. The lifetime of $D^{+}$ mesons was used to suppress combinatorial background by reconstructing the corresponding secondary vertex.
All data available after the HERA luminosity upgrade were used 
and this publication supersedes a previous one \cite{epj:c63:2009:2:171-188} which was based on a subset of the data. Three times more data were analysed for the current paper, with better control of the systematic uncertainties.

Differential cross sections were measured as a function of the photon virtuality at the electron vertex, $Q^2$, the inelasticity, $y$, and the transverse momentum,  $p_{T}(D^{+})$, and pseudorapidity, $\eta(D^{+})$, of the $D^{+}$ meson . 
The charm contribution to the proton structure-function $F_2$, denoted as \ftwocc, was extracted from the double-differential cross sections in $Q^2$ and $y$. 
Previous measurements, as well as NLO QCD predictions, are compared to the data.

\section{Experimental set-up}
\label{sec-exp}
The analysis was performed with data taken from 2004 to 2007, when HERA collided electrons or positrons with energy $E_{e} = 27.5\gev$ and protons with $E_{p} = 920\gev$, corresponding to a centre-of-mass energy $\sqrt{s} = 318\gev$. The corresponding integrated luminosity was $\Lumi = 354 \pm 7\pbi$.

\Zdetdesc

\Zctdmvddescnew\ZcoosysfnBetanew

\Zcaldescnew

\Zlumidescnew{1.9\%}

\section{Theoretical predictions}
\label{sec-theory}
Charm production in DIS has been calculated at NLO ($\ord{\als^2}$) in the  so-called fixed-flavour-number scheme (FFNS) \cite{np:b374:36}, in which the proton contains only light flavours and heavy quarks are produced in the hard interaction. 

The \hvqdis program \cite{pr:d57:2806} has been used to calculate QCD predictions for comparison  to the results of this analysis, as well as to extrapolate the measured visible cross sections to obtain \ftwocc. The renormalisation and factorisation scales were set to $\mu_R = \mu_F = \sqrt{Q^2+4 m_c^2}$ and the charm-quark pole mass to $m_c = 1.5\gev$. The FFNS variant of the ZEUS-S NLO QCD PDF fit \cite{pr:d67:012007} to inclusive structure-function data was used as the parametrisation of the proton PDFs. The same charm mass and choice of scales was used in the fit as in the HVQDIS calculation. The value of $\alsmzone$ in the three-flavour FFNS was set to $0.105$, corresponding to $\alsmzone = 0.116$ in the five-flavour scheme. 

To calculate $D^{+}$ observables, events at the parton level were interfaced with a fragmentation model based on the Kartvelishvili function \cite{pl:b78:615}.
The fragmentation was performed in the $\gamma^{*}p$ centre-of-mass frame.
The Kartvelishvili parameter, $\alpha$, was parametrised \cite{thesis:lisovyi:2011} as a smooth function of the invariant mass of the $c \bar{c}$ system, $M_{c \bar{c}}$, to fit the measurements of the $D^{*}$ fragmentation function by ZEUS \cite{jhep:04:082} and H1 \cite{epj:c59:589}: $\alpha (M_{c \bar{c}}) = 2.1 + 127/(M_{c \bar{c}}^2 - 4 m_c^2)$ (with $m_c$ and $M_{c \bar{c}}$ in $\Gev$). In addition, the mean value of the fragmentation function was scaled down by 0.95 since kinematic considerations \cite{jhep:0604:006} and direct measurements at Belle \cite{pr:d73:032002} and CLEO \cite{pr:d70:112001} show that, on average, the momentum of $D^{+}$ mesons is 5\% lower than that of $D^{*}$ mesons. This is due to some of the $D^{+}$ mesons originating from $D^{*}$ decays. For the hadronisation fraction, $f(c \rightarrow D^{+})$, the value $0.2297 \pm 0.0078$ was used \cite{upub:lohrmann:cfrac_arxiv}, having corrected  the branching ratios to those in the PDG 2012 \pcite{pr:d86:010001}.

The uncertainties on the theoretical predictions were estimated as follows:
 \begin{itemize}
	\item the renormalisation and factorisation scales were independently varied up and down by a factor 2;
	\item the charm-quark mass was consistently changed in the PDF fits and in the HVQDIS calculations by $\pm 0.15\gev$;
	\item the proton PDFs were varied within the total uncertainties of the ZEUS-S PDF fit;
	\item the fragmentation function was varied by changing the functional dependence of the parametrisation function $\alpha (M_{c \bar{c}})$ within uncertainties \cite{thesis:lisovyi:2011};
	\item the hadronisation fraction was varied within its uncertainties.
\end{itemize}
The total theoretical uncertainty was obtained by summing in quadrature the effects of the individual variations. 
The dominant contributions originate from the variations of the charm-quark mass and the scales. 
In previous studies~\cite{tech:desy12-XXX:ccombination} the uncertainty due to the variation of $\alsmzone$ was found to be insignificant and therefore it was neglected here.

A second NLO calculation was used in this analysis.
It is based on the general-mass variable-flavour-number scheme (GM-VFNS)~\cite{pr:d57:6871}.
In this scheme, charm production is treated in the FFNS in the low-$Q^2$ region, where the mass effects are largest, 
and in the zero-mass variable-flavour-number scheme (ZM-VFNS)~\cite{np:b278:934} at high $Q^2$. 
In the ZM-VFNS, the charm-quark 
mass is set to zero in the computation of the matrix elements and the kinematics.
Charm is treated as an active flavour in the proton above the kinematic threshold, $Q^2 \approx m_c^2$.
At intermediate scales, an interpolation is made in the GM-VFNS
between the FFNS and the ZM-VFNS, avoiding double counting of common terms. 

\section{Monte Carlo samples}
\label{sec-mc}

Monte Carlo (MC) simulations were used
to determine detector acceptances and to estimate and subtract
the contributions of $D^{+}$ mesons originating
from beauty decays. The MC events were generated with 
the \RAPGAP 3.00 \cite{cpc:86:147} program, interfaced with \HERACLES 4.6.1 \cite{cpc:69:155} to incorporate first-order electroweak corrections. The \RAPGAP generator uses the leading-order matrix element for the charm and beauty BGF process and parton showers 
to simulate higher-order QCD effects. The CTEQ5L \cite{epj:c12:375} PDFs were used for the proton. The heavy-quark masses were set to $m_c = 1.5\gev$ and $m_b = 4.75\gev$. The heavy-quark fragmentation was modelled using the Lund string model with the Bowler modifications for the longitudinal component \cite{zfp:c11:169}.

The generated events were passed through a full simulation of the ZEUS detector based on GEANT 3.21 \cite{tech:cern-dd-ee-84-1}. They had then to fulfil the same trigger criteria and pass the same reconstruction programs as the data.

\section{Selection of DIS events}
\label{sec-kinematics}

A three-level trigger system \cite{zeus:1993:bluebook, nim:a580:1257, *uproc:chep:1992:222} was used to select DIS events online. 
Most of the first-level triggers (FLTs) used in this analysis 
had some requirements on the track multiplicity in the events.
The efficiency of these criteria was measured using a trigger without track requirements and the detector simulation was tuned
to match the data.
The trigger-inefficiency corrections for the simulation were between $1\rnge10 \%$ for different tracking requirements.
The corrections changed the overall efficiency of the triggers used in the analysis \cite{thesis:lisovyi:2011} by a negligible 
amount for medium-$Q^2$ values and up to $\approx 2 \%$ for the low- and high-$Q^2$ regions.
At the third trigger level, DIS events were selected by requiring a scattered electron to be detected in the CAL. 

The kinematic variables $Q^2$ and $y$ were reconstructed offline using the double-angle (DA) method \cite{proc:hera:1991:23}, which relies on the angles of the scattered electron and the hadronic final state. To select a clean DIS sample the following cuts were applied:
\begin{itemize}
	\item $E_{e'} > 10\gev$, where $E_{e'}$ is the energy of the reconstructed scattered electron; 
	\item \sloppy $E^{\mathrm{cone}}_{\mathrm{non}\, e'} < 5\gev$, where $E^{\mathrm{cone}}_{\mathrm{non}\, e'}$ is the energy deposit in the CAL in a cone around the electron candidate, not originating from it. The cone was defined by the criterion \mbox{$\sqrt{\Delta\eta^2 + \Delta\phi^2} < 0.8$}, where $\phi$ is the azimuthal angle;\fussy
	\item for scattered electrons detected in the RCAL, the impact point of the candidate on the surface of the RCAL was required to lie outside a rectangular region ($\pm13\cm$ in $X$ and $\pm13\cm$ in $Y$) centred on the origin of coordinates;
	\item $40 < \delta < 65\gev$, where $\delta = \sum_i E_i (1 - \cos \theta_i)$ and $E_i$ and $\theta_i$ are the energy and the polar angle of the \emph{i}\textsuperscript{th} energy-flow object (EFO) \cite{thesis:briskin:1998, *epj:c1:81} reconstructed from CTD+MVD tracks and energy clusters measured in the CAL. This cut was imposed to select fully contained neutral-current $ep$ events, for which $\delta = 2 E_{e} = 55\gev$ and to suppress photoproduction contamination and cosmic-ray background;
	\item $|Z_{\mathrm{vtx}}| < 30\cm$, where $Z_{\mathrm{vtx}}$ is the $Z$ position of the primary vertex;
	\item $y_{\mathrm{JB}} > 0.02$, where $y$ was reconstructed using the Jacquet-Blondel method \cite{proc:epfacility:1979:391}.
\end{itemize}
The selected kinematic region was $5 < Q_\DA^2 < 1000\gev^2$ and $0.02 < y_\DA < 0.7$.

\section{Reconstruction of \boldmath{$D^{+}$} mesons}
\label{sec-dch}

The $D^{+}$ mesons were reconstructed using the decay channel $D^{+} \rightarrow K^{-}\pi^{+}\pi^{+}$. In each event, track pairs with equal charges were combined with a third track with the opposite charge to form $D^{+}$ candidates. The pion mass was assigned to the tracks with equal charges and the kaon mass was assigned to the remaining track. The three tracks were then fitted to a common vertex and the invariant mass, $M(K\pi\pi)$, was calculated. The tracks were required to have a transverse momentum $p_T(K) > 0.5\gev$ and $p_T(\pi) > 0.35\gev$, respectively. To ensure high momentum and position resolution, all tracks were required to be reconstructed within  $|\eta| < 1.75$, to  have passed through at least three superlayers of the CTD and to have at least two BMVD hits in the $XY$ plane and two in the $Z$ direction. 
A special study~\cite{private:libov} was performed to assess the tracking inefficiency for charged pions due to hadronic interactions in the detector material and how well the MC reproduces these interactions.
The MC simulation was found to underestimate the interaction rate by about $40\%$ for $p_T < 1.5\gev$
and to agree with the data for $p_T > 1.5\gev$.
A corresponding correction was applied to the MC.
The effect of the correction on the $D^{+}$ production cross section was about $3 \%$.

The kinematic region for $D^{+}$ candidates was \mbox{$1.5 < p_T(D^{+}) < 15\gev$} and \mbox{$|\eta(D^{+})| < 1.6$}. The contribution from \mbox{$D^{*+} \rightarrow D^0 \pi^{+} \rightarrow K^{-} \pi^{+} \pi^{+}$}  was suppressed by removing combinations with \mbox{$M(K\pi\pi) - M(K\pi) <0.15\gev$}. A small contribution from $D_s^{+} \rightarrow \phi \pi^{+} \rightarrow K^{-} K^{+} \pi^{+}$ was suppressed by assuming one of the pions to be a kaon and requiring the invariant mass of the kaon pair to lie outside the $\phi$ mass peak region $1.0115 < M(K K) < 1.0275\gev$. A remaining reflection 
from  $D_s^{+} \rightarrow K^{-} K^{+} \pi^{+}$ decays without intermediate
$\phi$ production 
 was estimated using the \RAPGAP MC sample. It was found to be $\approx 1\%$ and was subtracted from the mass distribution.

A powerful discriminating variable to suppress combinatorial background originating from light-flavour production is the decay-length significance, $S_l$. It is defined as $S_l = l/\sigma_l$, where $l$ is the decay length in the transverse plane, projected on to the $D^{+}$ meson momentum, and $\sigma_l$ is the uncertainty associated with this distance. 
The decay length itself was determined as the distance in  $XY$ 
between the secondary vertex fitted in 3D and the interaction point. In the $XY$ plane, the interaction point is defined as the position of the primary vertex determined from selected tracks and using the beam-spot position \cite{epj:c63:2009:2:171-188}
as an additional constraint. The widths of the beam spot were $88\mum$ ($80\mum$) and $24\mum$ ($22\mum$) in the $X$ and $Y$ directions, respectively, for the $e^{+}p$ ($e^{-}p$) data. 

Only candidates with a decay length in the $XY$ plane less than $1.5\cm$ were selected in the analysis to ensure that the vertex was inside the beam pipe, suppressing background caused by interactions in the beam pipe or detector material. The $S_l$ distribution is asymmetric with respect to zero, with charm mesons dominating in the positive tail. Detector resolution effects cause the negative tail, which is dominated by light-flavour events. A smearing \cite{thesis:lisovyi:2011, thesis:schoenberg:2010} was applied to the decay length of a small fraction of the MC events in order to reproduce the negative decay-length data. 
Finally, a cut \mbox{$S_l > 4$} was applied; according to MC studies this optimises the statistical precision of the measurement.
In addition, the $\chi^2$ of the fitted secondary vertex, $\chi^2_{\mathrm{sec.vtx.}}$, was required to be less than 10 for three degrees of freedom, to ensure good quality of the reconstructed $D^{+}$ vertex. 

\begin{figure}[p]
	\begin{center}
	\includegraphics[width=0.95\textwidth]{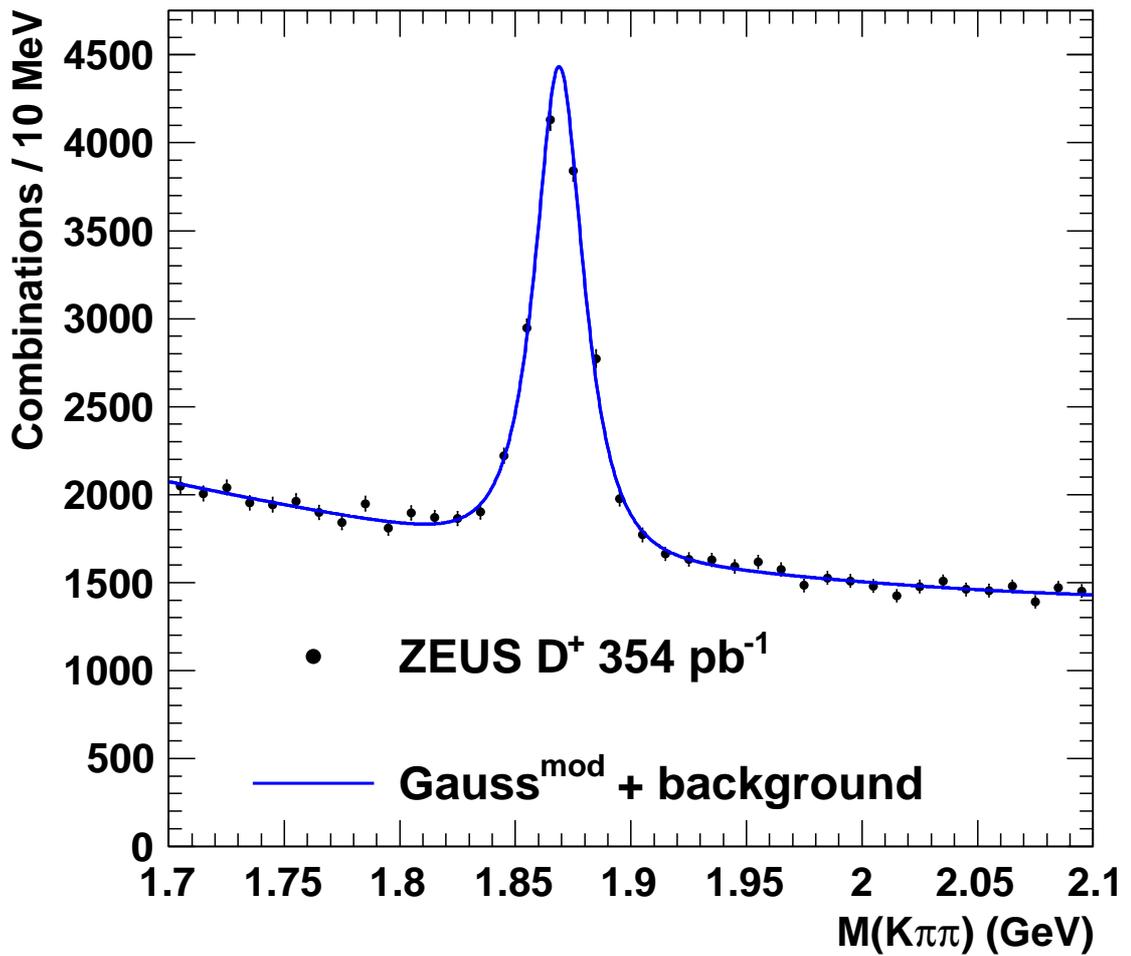}
	\end{center}
	\caption{
	Mass distribution of the reconstructed $D^{+}$ candidates. The solid curve represents a fit by the sum of a modified Gaussian for the signal and a second-order polynomial for the background.}
	\label{fig-mass}
\end{figure}

\Fig{mass} shows the $M(K\pi\pi)$ distribution for the selected $D^{+}$ candidates. To extract the number of reconstructed $D^{+}$ mesons, the mass distribution was fitted with the sum of a modified Gaussian function for the signal and a second-order polynomial to parametrise the background. The fit function was integrated over each bin. The modified Gaussian function had the form:
\begin{equation*}
        \mathrm{Gauss^{mod}} \propto \mathrm{exp}[-0.5 \cdot x^{1+1/(1+\beta\cdot x)}],
\end{equation*}
where $x = |(M(K\pi\pi) - M_{0})/\sigma|$ and $\beta = 0.5$. This functional form describes the signals in both the data and MC simulations well. The signal position, $M_{0}$, the width, $\sigma$, as well as the number of $D^{+}$ mesons were free parameters in the fit. The number of $D^{+}$ mesons yielded by the fit was $N(D^{+}) = 8356 \pm 198$. The fitted position of the peak was $M_{0} = 1868.97 \pm 0.26\mev$, where only the statistical uncertainty is quoted, 
consistent with the PDG value of $1869.62 \pm 0.15\mev$~\cite{pr:d86:010001}.


\section{Cross-section determination}
\label{sec-acceptance}

For a given observable, $Y$, the differential cross section in the \emph{i}\textsuperscript{th} bin was determined as 
\begin{equation*}
        \frac{d \sigma}{d Y} = \frac{N^{i} - N^{i}_{b}}{\mathcal{A}_{c}^{i} \, \mathcal{L} \, \mathcal{B} \Delta Y^{i}} \cdot \mathcal{C}_{\rmrad}^{i} , 
\end{equation*}
where $N^{i}$ is the number of reconstructed $D^{+}$ mesons in bin \emph{i} of size $\Delta Y^{i}$. The reconstruction acceptance, $\mathcal{A}_{c}^{i}$, takes into account geometrical acceptances, detector efficiencies and migrations due to the finite detector resolution. The values of $\mathcal{A}_{c}^{i}$ were determined using the \RAPGAP MC simulation for charm production in DIS (see \Sect{mc}).
The quantity $\mathcal{L}$ denotes the integrated luminosity and $\mathcal{B}$  the branching ratio
for the $D^{+} \rightarrow K^{-}\pi^{+}\pi^{+}$ decay channel, which is $9.13 \pm 0.19 \%$~\cite{pr:d86:010001}. 
The radiative corrections, $\mathcal{C}_{\rmrad}^{i}$, were used to correct measured cross sections to the Born level.
For the acceptance determination, the charm MC events were reweighted \cite{thesis:lisovyi:2011} to reproduce the $Q^2$, $p_{T}(D^{+})$ and $\eta(D^{+})$ distributions in the data.

For all measured cross sections, the contribution of reconstructed $D^{+}$ mesons originating from beauty production, $N^{i}_{b}$, was subtracted using the prediction from the \RAPGAP MC simulation.
This prediction was scaled by a factor $1.6$, an average value which was estimated from previous ZEUS measurements \citeBeauty of beauty production in DIS. The subtraction of the \emph{b}-quark contribution reduced the measured cross sections by $5 \%$ on average.

The measured cross sections were corrected to the QED Born level, calculated using a running coupling constant, $\alpha$, such that they can be directly compared to the QCD predictions by \HVQDIS. The \RAPGAP Monte Carlo was used to calculate $\mathcal{C}_{\rmrad} = \sigma_{\rmBorn} / \sigma_{\rmrad}$, where $\sigma_{\rmrad}$ is the predicted cross section with full QED corrections (as in the default MC samples) and $\sigma_{\rmBorn}$ was obtained with QED corrections turned off, keeping $\alpha$ running. The corrections are typically $\mathcal{C}_{\rmrad} \approx 1.02$ and reach $1.10$ in the high-$Q^2$ region.

\begin{figure}[p]
        \begin{center}        
        \includegraphics[width=0.95\textwidth]{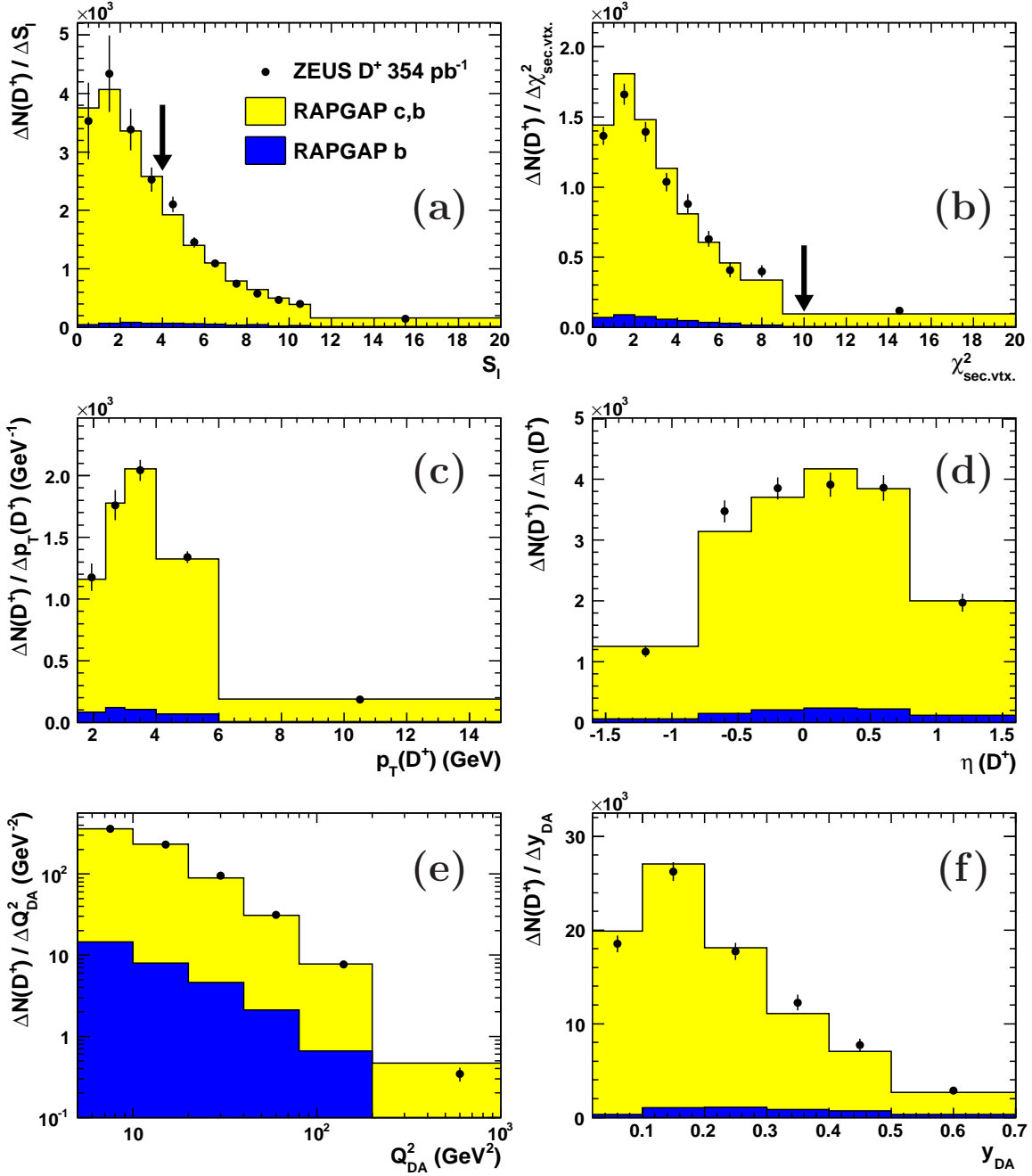}			\Text(-250,410)[]{\Large{\textbf{(a)}}} \Text(-30,410)[]{\Large{\textbf{(b)}}} \Text(-250,298)[]{\Large{\textbf{(c)}}} \Text(-30,298)[]{\Large{\textbf{(d)}}} \Text(-250,130)[]{\Large{\textbf{(e)}}} \Text(-30,130)[]{\Large{\textbf{(f)}}}
	\end{center}
        \caption{Bin-averaged differential $D^{+}$ distributions of (a) $S_l$ , (b) $\chi^2_{\mathrm{sec.vtx.}}$, (c) $p_{T}(D^{+})$, (d) $\eta(D^{+})$, (e) $Q^2_\DA$, (f) $y_\DA$. The $S_l$ and $\chi^2_{\mathrm{sec.vtx.}}$ distributions are shown before the final selection cuts indicated by vertical arrows. The data are shown as black points, with bars representing the statistical uncertainty. Also shown are the simulated charm+beauty MC distributions (light shaded area). The beauty contribution (dark shaded area) is shown separately. The sum of the charm+beauty MC simulations was normalised to the data area.}
        \label{fig-cp}
\end{figure}

\Fig{cp} shows important variables for the secondary-vertex reconstruction, distributions for the DIS variables and the kinematics of the $D^{+}$ meson. For all variables, the number of reconstructed $D^{+}$ mesons was extracted fitting the number of $D^{+}$ mesons in each bin of the distribution.
The reweighted MC provides a reasonable description of the data.


\section{Systematic uncertainties}
\label{sec-syst}

The systematic uncertainties were determined by changing the analysis procedure or varying parameter values within their estimated uncertainties and repeating the extraction of the signals and the cross-section calculations. The following sources of systematic uncertainties were considered with the typical effect on the cross sections given in parentheses:
\begin{itemize}
	\item $\{\delta_1\}$ the cut on the positions $|X|$ and $|Y|$ of the scattered electron in the RCAL was varied by $\pm 1\cm$ in both the data and the MC simulations, to account for potential imperfections of the detector simulation near the inner edge of the CAL ($\pm 1\%$);
	\item $\{\delta_2\}$ the reconstructed electron energy was varied by $\pm 2 \%$ in the MC only, to account for the uncertainty in the electromagnetic energy scale ($< 1\%$);
	\item $\{\delta_3\}$ the energy of the hadronic system was varied by $\pm 3 \%$ in the MC only, to account for the uncertainty in the hadronic energy scale ($< 1\%$);
	\item $\{\delta_4\}$ the FLT tracking-efficiency corrections for the MC (see \Sect{kinematics}) were varied within the estimated uncertainties associated to them ($< 1\%$);
	\item uncertainties due to the signal-extraction procedure were estimated repeating the fit in both the data and the MC using:
	\begin{itemize}
		\item[-] $\{\delta_5\}$ an exponential function for the background parametrisation ($< 1\%$);
		\item[-] $\{\delta_6\}$ a signal parametrisation changed by simultaneously varying the $\beta$ parameter of the modified Gaussian function in the data and MC by ${}^{+0.1}_{-0.2}$ from the nominal value $0.5$. The range was chosen to cover the values which give the best description of the mass peaks in the data and MC simulations in bins of the differential cross sections (${}^{+0.7\%}_{-1.5\%}$);
	\end{itemize}
	\item $\{\delta_7\}$ the effect of the decay-length smearing procedure was varied by $\pm50\%$ of its size, to estimate the uncertainty due to the decay-length description ($\pm 1\%$). As a further cross check, the cut on the decay-length significance was varied between 3 and 5. The resulting variations of the cross sections were compatible with the variation of the decay-length smearing and were therefore omitted to avoid double counting; 
	\item $\{\delta_8\}$ the scaling factor for the MC beauty-production cross sections was varied by $\pm 0.6$ from the nominal value $1.6$. This was done to account for the range of the \RAPGAP beauty-prediction normalisation factors extracted in various analyses \citeBeauty ($\pm 2\%$);
	\item the uncertainties due to the model dependence of the acceptance corrections were estimated by varying the shapes of the kinematic distributions in the charm MC sample in a range of good description of the data \cite{thesis:lisovyi:2011}:
	\begin{itemize}
		\item[-] $\{\delta_9\}$ the $\eta(D^{+})$ reweighting function was varied ($\pm 2\%$);
		\item[-] $\{\delta_{10}\}$ the shapes of the $Q^2$ and $p_{T}(D^{+})$ were varied simultaneously ($\pm 4\%$);
	\end{itemize}	
	\item $\{\delta_{11}\}$ the uncertainty 
of the pion track inefficiency due to nuclear interactions (see \Sect{dch}) was evaluated by 
varying the correction applied to the MC by its estimated uncertainty of $\pm 50 \%$ of its nominal size ($\pm 1.5\%$);
	\item overall normalisation uncertainties:
	\begin{itemize}
		\item[-] $\{\delta_{12}\}$ the simulation of the MVD hit efficiency ($\pm 0.9 \%$);	
		\item[-] $\{\delta_{13}\}$ the effect of the imperfect description of $\chi^2_{\mathrm{sec.vtx.}}$ was checked by multiplying $\chi^2_{\mathrm{sec.vtx.}}$ for $D^{+}$ candidates in the MC simulations by a factor 1.1 to match the distribution in the data ($+ 2\%$);	
		\item[-] $\{\delta_{14}\}$ the branching ratio uncertainty ($\pm 2.1 \%$);
		\item[-] $\{\delta_{15}\}$ the measurement of the luminosity ($\pm 1.9 \%$).
	\end{itemize}	
\end{itemize}
The size of each systematic effect was estimated bin-by-bin except for the normalisation uncertainties ($\delta_{12} \rnge \delta_{15}$). The overall systematic uncertainty was determined by adding the above uncertainties in quadrature. The normalisation uncertainties due to the luminosity measurement and that of the branching ratio were not included in the systematic uncertainties on the differential cross sections. 

\section{Results}
\subsection{Cross sections}
\label{sec-xs}

The production of $D^{+}$ mesons in the process $e p \rightarrow e' c \bar{c} X \rightarrow e' D^{+} X'$ (i.e.\ not including $D^{+}$ mesons from beauty decays) was measured in the kinematic range:
\begin{equation*}
        5 < Q^{2} < 1000\gev^{2}, \, 0.02 < y < 0.7, \, 1.5 < p_{T}(D^{+}) < 15\gev, \, |\eta(D^{+})| < 1.6.
\end{equation*}

\begin{figure}[p]
	\begin{center}
	\includegraphics[width=0.80\textwidth]{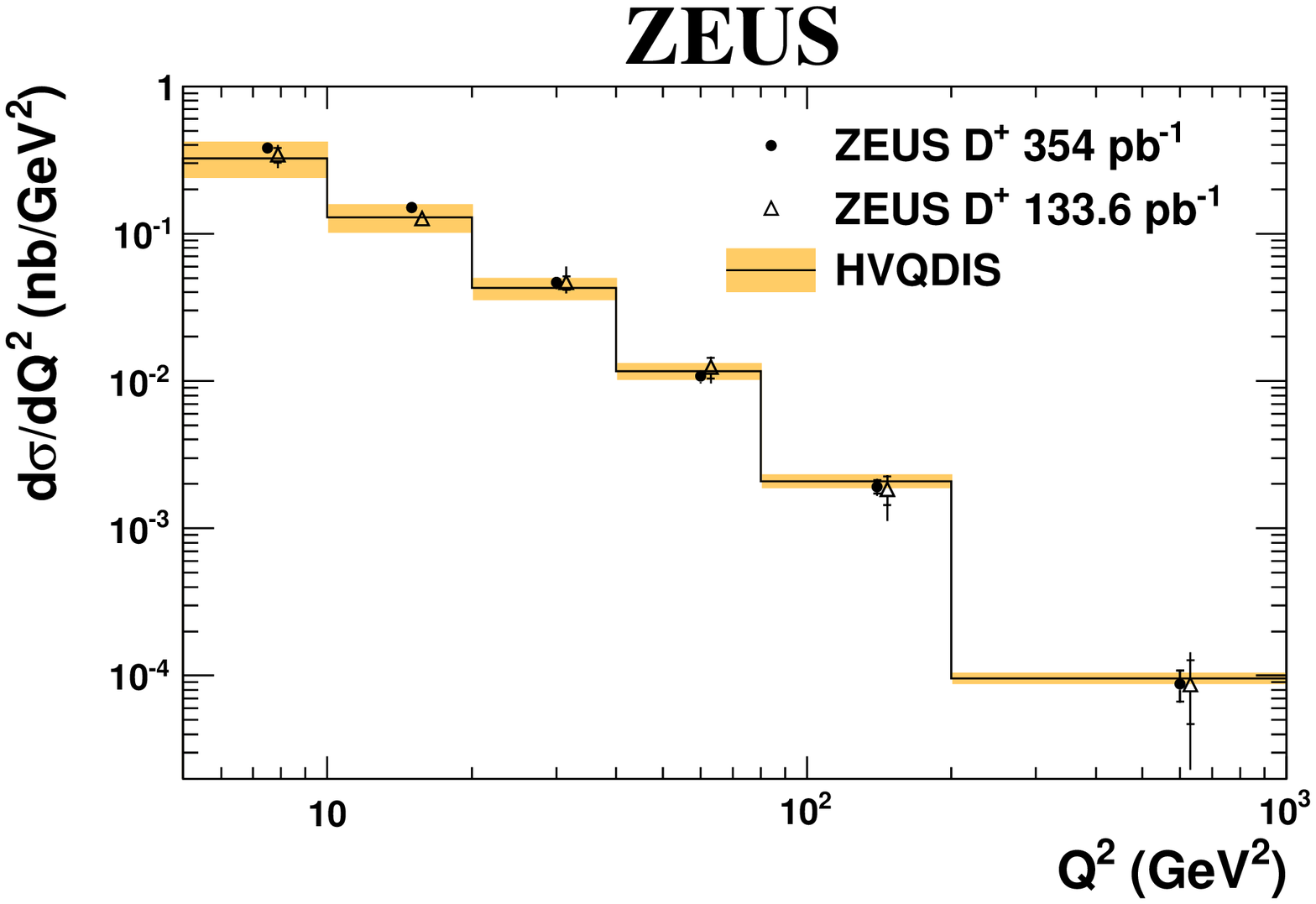} \Text(-60,140)[]{\Large{\textbf{(a)}}}
	\includegraphics[width=0.80\textwidth]{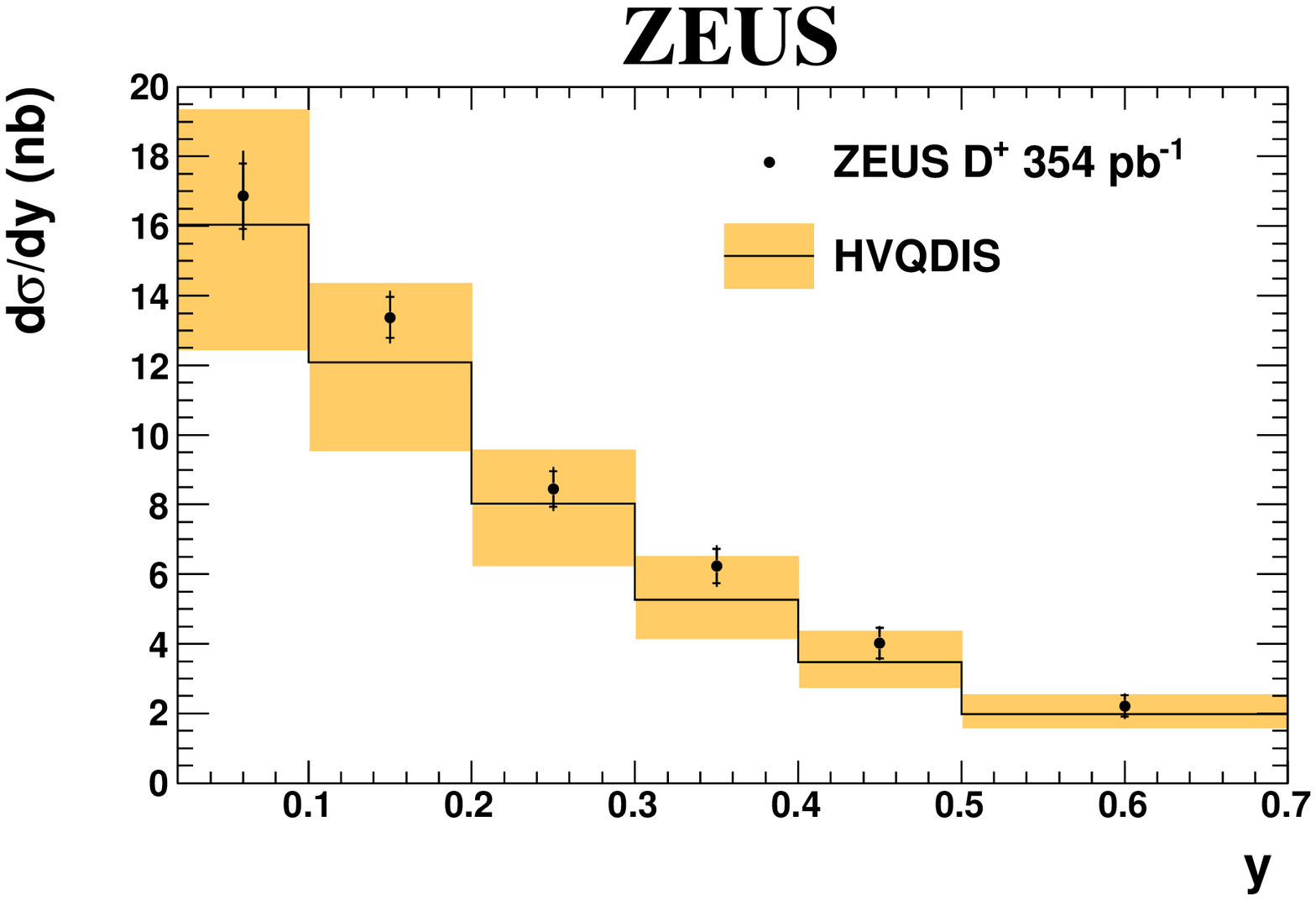}  \Text(-60,140)[]{\Large{\textbf{(b)}}}
	\end{center}
	\caption{Bin-averaged differential cross sections for $D^{+}$ meson production in the process $e p \rightarrow e' c \bar{c} X \rightarrow e' D^{+} X'$ as a function of (a) $Q^{2}$ and (b) $y$. The cross sections are given in the kinematic region \mbox{$5 < Q^{2} < 1000\gev^{2}$}, \mbox{$0.02 < y < 0.7$}, \mbox{$1.5 < p_{T}(D^{+}) < 15\gev$} and \mbox{$|\eta(D^{+})| < 1.6$}. The results obtained in this analysis  are shown as filled circles. The inner error bars correspond to the statistical uncertainty, while the outer error bars represent the statistical and systematic uncertainties added in quadrature. For the cross section as a function of $\,Q^{2}$, the results of the previous ZEUS measurement are also shown (open triangles). The solid lines and the shaded bands represent the NLO QCD predictions in the FFNS with estimated uncertainties.}
	 \label{fig-xs_sd2}
\end{figure}

The differential cross sections as a function of $Q^{2}$ and $y$ are shown in \fig{xs_sd2}. The cross section falls by about three orders of magnitude over the measured $Q^2$ range and one order of magnitude in $y$. The data presented here are in good agreement with the previous ZEUS $D^+$ measurement\footnote{The contribution of $D^{+}$ mesons from beauty decays was subtracted using the scaled \RAPGAP MC predictions.}~\cite{epj:c63:2009:2:171-188}. They have significantly smaller uncertainties and supersede the previous results. The NLO QCD predictions calculated in the FFNS, using HVQDIS~\cite{pr:d67:012007}, provide a good description of the measurements. The experimental uncertainties are smaller than the theoretical uncertainties, apart from the high-$Q^2$ region, where statistics is limited.

\begin{figure}[p]
	\begin{center}
	\includegraphics[width=0.85\textwidth]{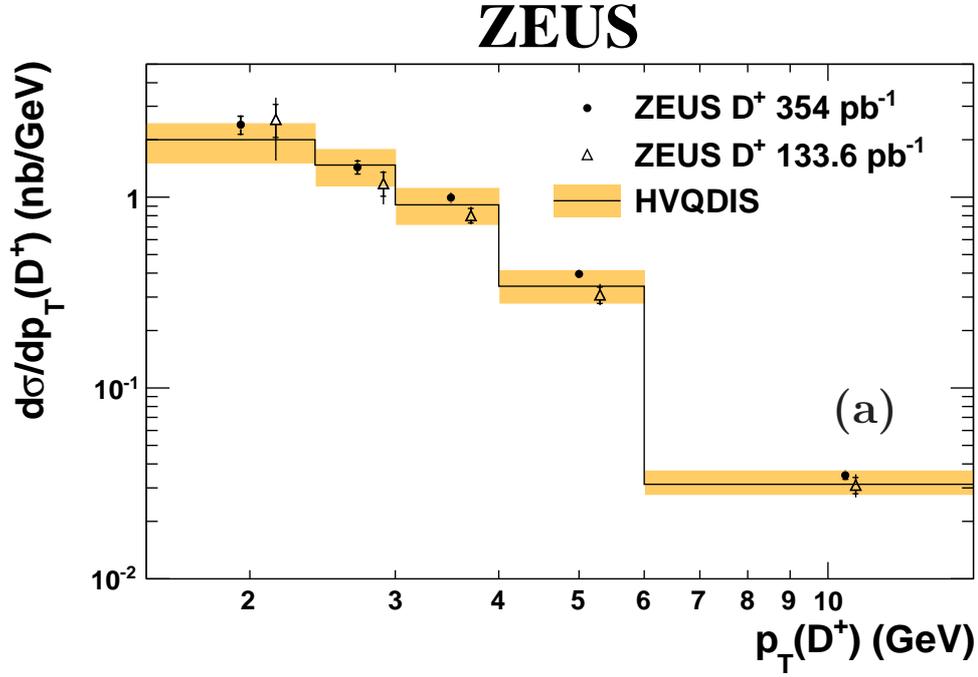}  \Text(-60,100)[]{\Large{\textbf{(a)}}}
	\includegraphics[width=0.85\textwidth]{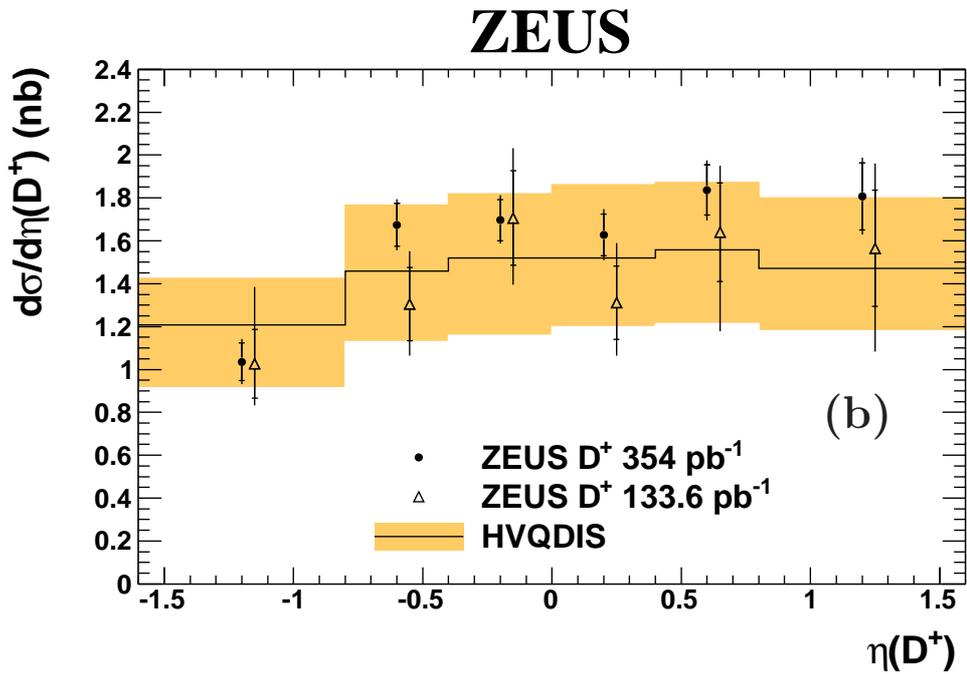} \Text(-60,100)[]{\Large{\textbf{(b)}}}
	\end{center}
	\caption{Bin-averaged differential cross sections for $D^{+}$ meson production in the process $e p \rightarrow e' c \bar{c} X \rightarrow e' D^{+} X'$ as a function of (a) $p_{T}(D^{+})$ and (b) $\eta(D^{+})$. Other details are as in  \protect \fig{xs_sd2}.}
	 \label{fig-xs_sd1}
\end{figure}

\Fig{xs_sd1} shows that the $D^+$ cross section also falls with the transverse momentum, $p_{T}(D^{+})$, but is only mildly dependent on the pseudorapidity, $\eta(D^{+})$. The HVQDIS calculation describes the behaviour of the data well. The results shown in \figand{xs_sd2}{xs_sd1} are listed in \taband{xs_sd2}{xs_sd1}.

\begin{figure}[p]
	\begin{center}
	\includegraphics[width=0.43\textwidth]{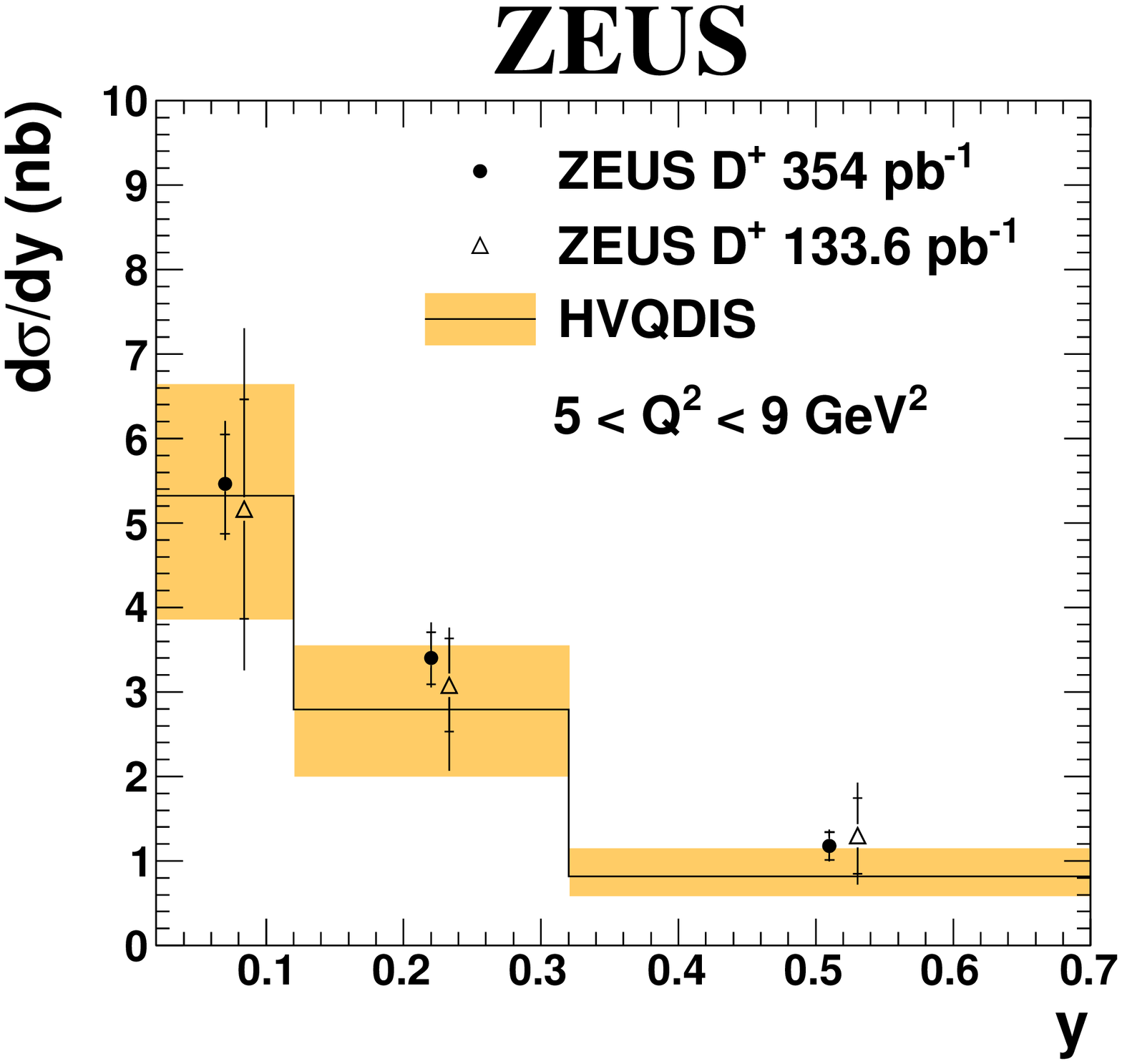} \Text(-40,85)[]{\large{\textbf{(a)}}}
	\includegraphics[width=0.43\textwidth]{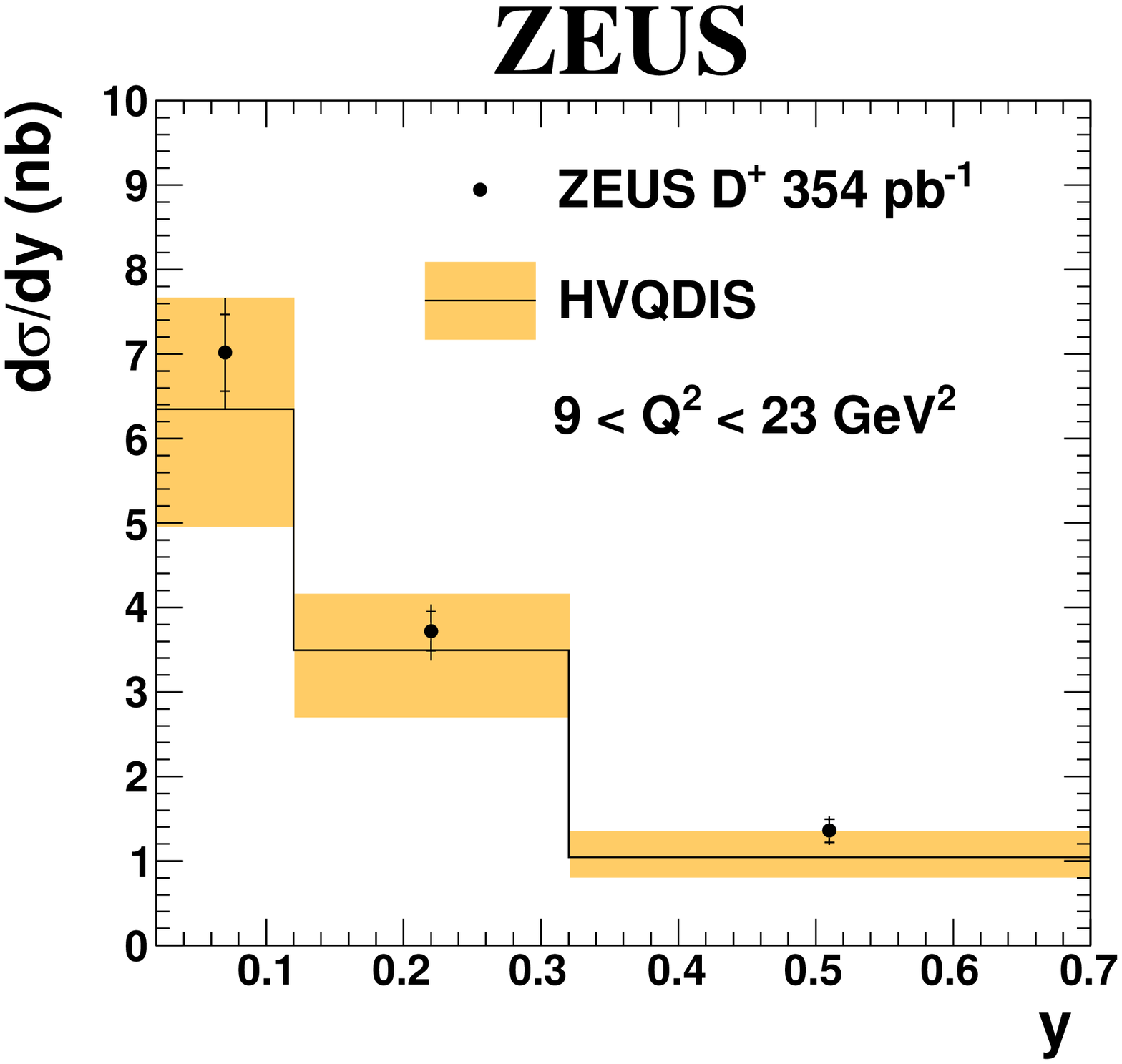} \Text(-40,85)[]{\large{\textbf{(b)}}}
        \includegraphics[width=0.43\textwidth]{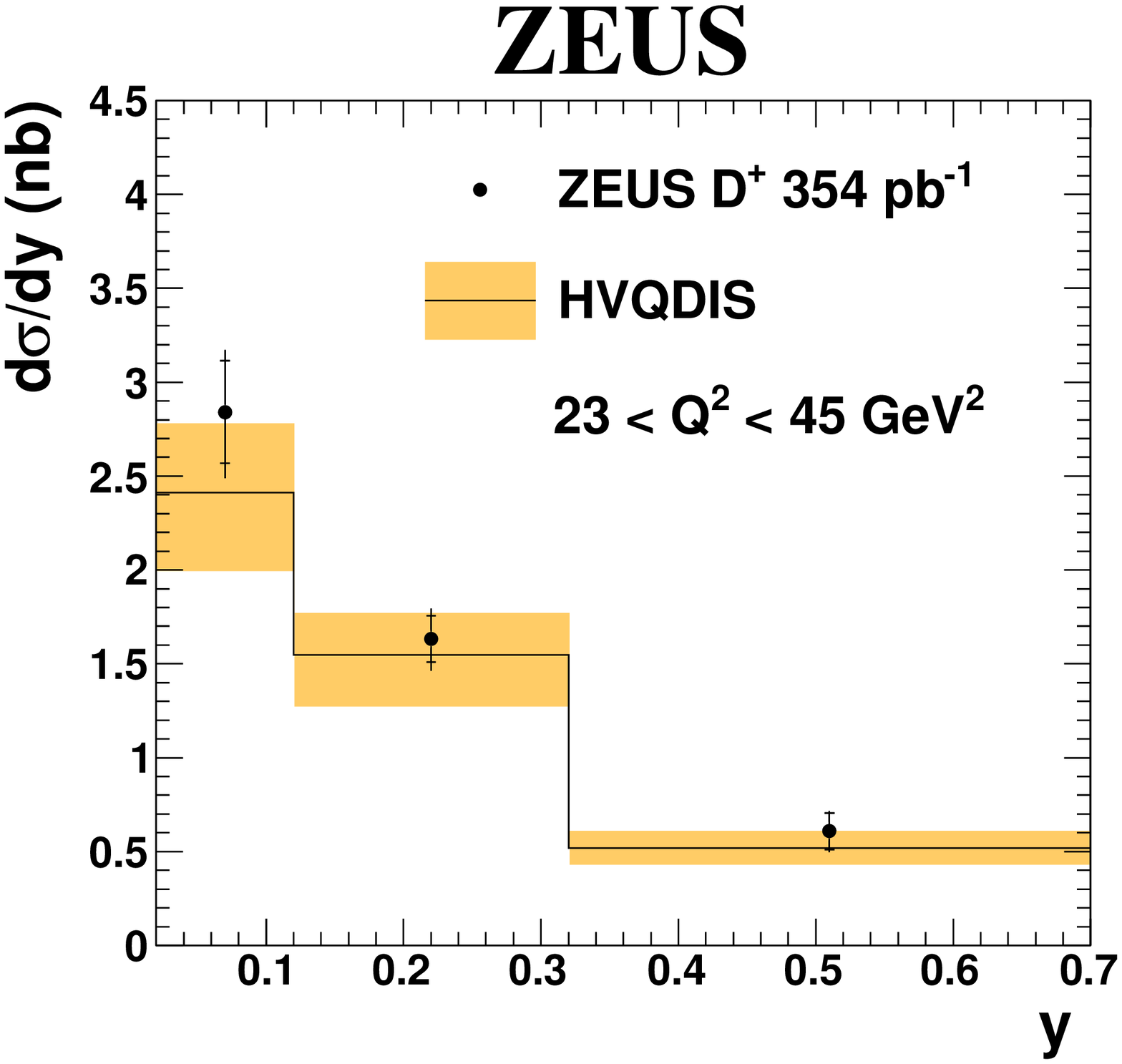} \Text(-40,85)[]{\large{\textbf{(c)}}}
	\includegraphics[width=0.43\textwidth]{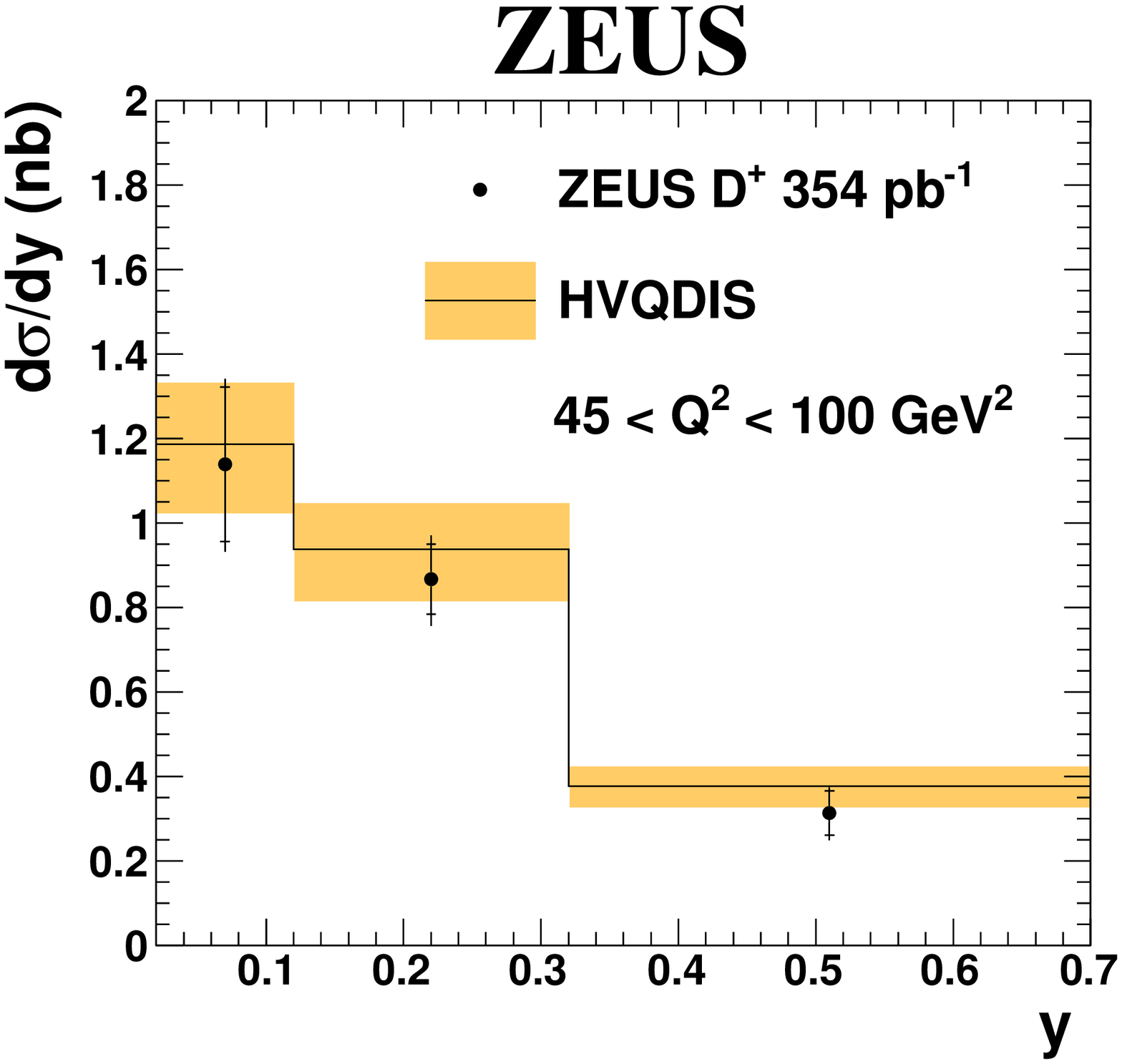} \Text(-40,85)[]{\large{\textbf{(d)}}}
	\includegraphics[width=0.43\textwidth]{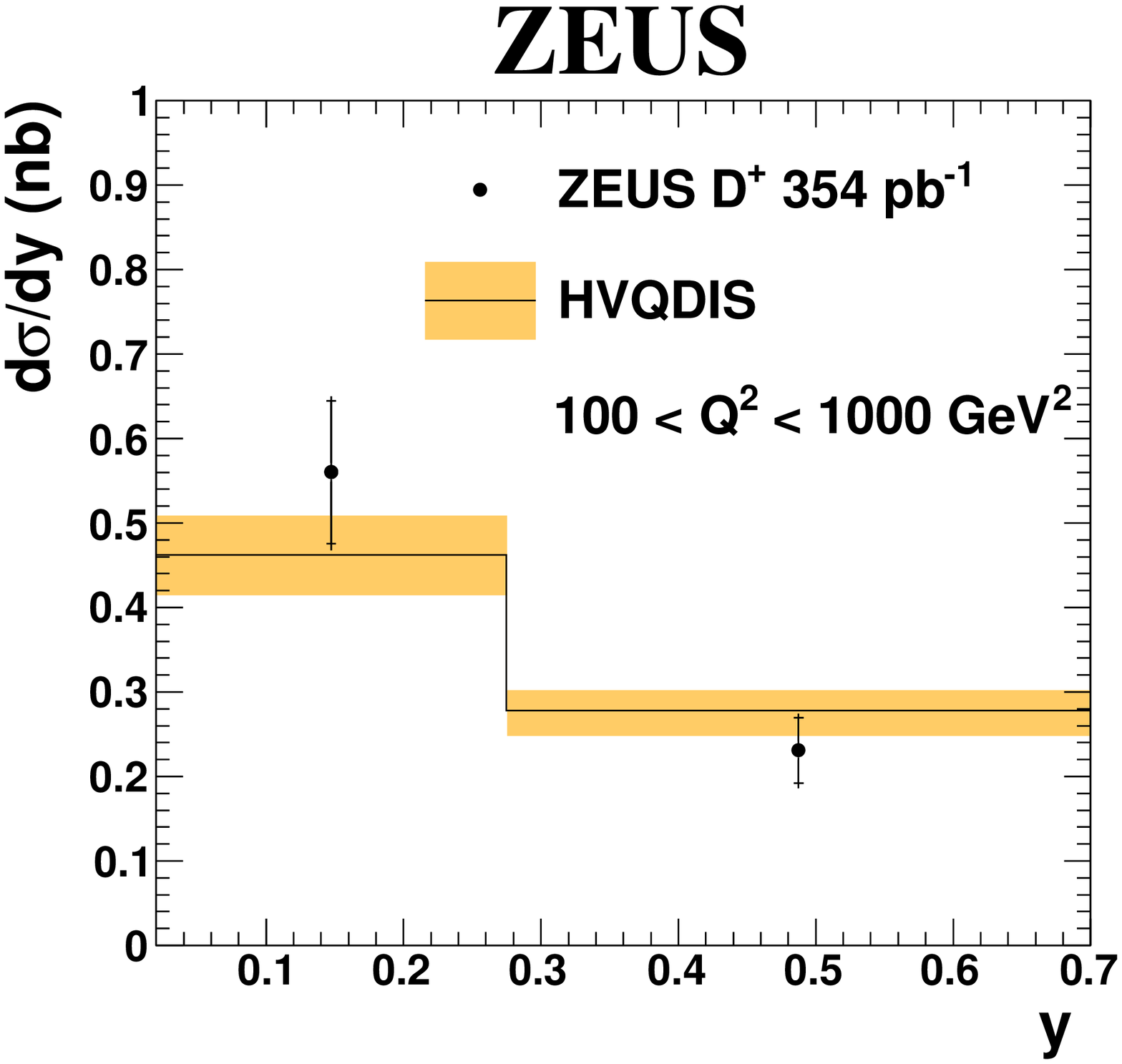} \Text(-40,85)[]{\large{\textbf{(e)}}}
	\end{center}
	\caption{Bin-averaged differential cross sections for $D^{+}$ meson production in the process $e p \rightarrow e' c \bar{c} X \rightarrow e' D^{+} X'$ as a function of $y$ in different $Q^2$ ranges: (a)~\mbox{$5 < Q^{2} < 9\gev^{2}$}, (b)~\mbox{$9 < Q^{2} < 23\gev^{2}$}, (c)~\mbox{$23 < Q^{2} < 45\gev^{2}$}, (d)~\mbox{$45 < Q^{2} < 100\gev^{2}$} and (e)~\mbox{$100 < Q^{2} < 1000\gev^{2}$}. Other details are as in  \protect \fig{xs_sd2}.}
	 \label{fig-xs_dd}
\end{figure}

\Fig{xs_dd} shows the differential cross sections as a function of $y$ in five $Q^2$ ranges. The data are well reproduced by the HVQDIS calculation. The cross-section values are given in \tab{xs_dd}. The effects of individual sources of systematic uncertainties (described in \Sect{syst}) on the cross sections in bins of $Q^2$ and $y$ are given in \tab{syst_dd}. 

\subsection{Extraction of \boldmath{\ftwocc}}
\label{sec-f2c}

The inclusive double-differential $c \bar{c}$ cross section in $Q^2$ and $x = Q^2/s y$ can be expressed as
\begin{equation*}
        \frac{d\sigma^{c\bar{c}}}{dx \, dQ^2} = \frac{2 \pi \alpha^2}{x \, Q^4} \Big{[} (1+(1-y)^2) \, F_2^{c\bar{c}} - y^2 \, F_L^{c\bar{c}} \Big{]},
	\label{eq-structfunct_c}
\end{equation*}
where \ftwocc and $F_L^{c\bar{c}}$ denote the charm contributions to the structure-function $F_2$ and the longitudinal structure function, $F_L$, respectively.

The differential $D^{+}$ cross sections, $\sigma_{i, \mathrm{meas}}$, measured in bins of $Q^2$ and $y$ (\tab{xs_dd}), were used to extract \ftwocc at reference points $Q^2_{i}$ and $x_{i}$ within each bin, using the relationship
\begin{equation}
        F_{2, \mathrm{meas}}^{c\bar{c}} (x_{i},Q^{2}_{i}) = \sigma_{i, \mathrm{meas}} \frac{F_{2, \mathrm{theo}}^{c\bar{c}} (x_{i},Q^{2}_{i})}{\sigma_{i, \mathrm{theo}}},
	\label{eq-F2c_extrapolation}
\end{equation}
where $F_{2, \mathrm{theo}}^{c\bar{c}}$ and $\sigma_{i, \mathrm{theo}}$ were calculated at NLO in the FFNS using the \HVQDIS program. 
This procedure corrects for 
the $f(c \rightarrow D^{+})$ hadronisation fraction and for 
the extrapolation from the restricted kinematic region of the $D^{+}$ measurement (\mbox{$1.5 < p_{T}(D^{+}) < 15\gev$}, $|\eta(D^{+})| < 1.6$) to the full phase space.
The extrapolation factors were found to vary 
from $1.5$ at high $Q^2$ to $3.0$ at low $Q^2$. The uncertainty on the extrapolation procedure was estimated 
by applying the same variations that were used to determine the uncertainty of the HVQDIS theoretical predictions (see \Sect{theory})
for the ratio $F_{2, \mathrm{theo}}^{c\bar{c}} (x_{i},Q^{2}_{i}) / \sigma_{i, \mathrm{theo}}$ and adding the resulting ratio uncertainties in quadrature.
The procedure of Eq.~\eq{F2c_extrapolation} also corrects for the $F_L$ contribution to the cross section.
This assumes that the \HVQDIS calculation correctly predicts the ratio $F_L^{c\bar{c}}/F_{2}^{c\bar{c}}$.
This calculation yields a contribution of $F_L^{c\bar{c}}$ between $0\%$ and $3\%$ at low and high $y$, respectively.

\begin{figure}[p]
	\begin{center}
	\includegraphics[width=0.97\textwidth]{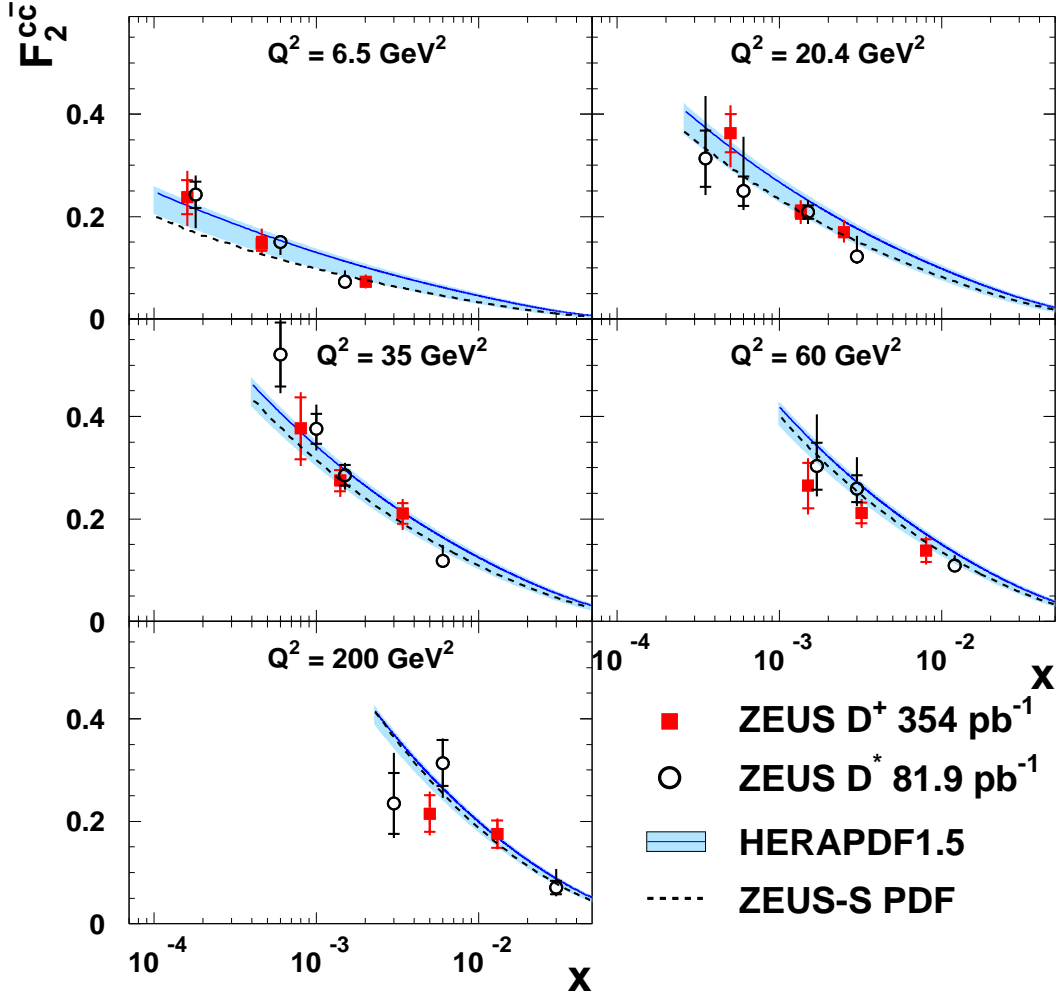}
	\end{center}
	\caption{Structure-function \ftwocc as a function of $x$ for various values of $Q^2$. Results obtained in this analysis are shown as filled squares. Also shown are the results of a previous \ftwocc measurement by ZEUS (open points) based on $D^{*}$ production. The inner error bars correspond to the statistical uncertainty, while the outer error bars represent the statistical, systematic and theoretical uncertainties added in quadrature. Also shown are predictions in the GM-VFNS based on HERAPDF1.5 with the charm-quark-mass parameter set to $1.4\gev$ for the central value (solid line) and its variation in the range $1.35\gev$ to $1.65\gev$ (filled band). Predictions in the FFNS based on the ZEUS-S PDF set with the default settings described in the text are shown as well (dashed line).}
	 \label{fig-f2c}
\end{figure}

The extracted values of \ftwocc are presented in \tab{f2c} and \fig{f2c}. 
\Fig{f2c} also shows a comparison to a previous ZEUS measurement of \ftwocc using $D^{*}$ mesons \cite{pr:d69:012004}. The previous results were corrected to the $Q^2$ grid used in the present analysis using NLO QCD calculations. The two measurements are in good agreement and have similar precision. NLO QCD predictions in the FFNS and GM-VFNS were also compared to the data. The FFNS predictions correspond to the calculations that were used in the \ftwocc extraction. The GM-VFNS calculations are based on the HERAPDF1.5 \cite{misc:ichep10:168} PDF set with the charm-quark-mass parameter set to $1.4\gev$. The band shows the result of the variation of the charm-quark-mass parameter from $1.35\gev$ to $1.65\gev$ in the calculations. Both predictions provide a good description of the data.

\subsection{Reduced cross section}

The results on $D^+$ presented here can be combined with other measurements on charm production. 
Their systematics are largely independent of those using other tagging methods. 
In such combinations, the quantity used is the reduced charm cross section, defined as 

\begin{equation*}
       \sigma^{c\bar{c}}_{\mathrm{red}} = \frac{d\sigma^{c\bar{c}}}{dx \, dQ^2} \cdot \frac{x \, Q^4}{2 \pi \alpha^2 \, (1+(1-y)^2)} = F_2^{c\bar{c}} - \frac{y^2}{1+(1-y)^2} \, F_L^{c\bar{c}}.
	\label{eq-sigma_c_red}
\end{equation*}
The extraction of $\sigma^{c\bar{c}}_{\mathrm{red}}$ closely follows the determination
of $F_2^{c\bar{c}}\,$; a modified version of Eq.~\eq{F2c_extrapolation} is used,  
simply replacing on both sides the structure function by the reduced cross section.
The reduced cross sections of the present analysis  are corrected to the same $Q^2$ values as in the charm combination paper of H1 and ZEUS \cite{tech:desy12-XXX:ccombination} 
and are presented in \tab{sigmac}.


\section{Conclusions}
\label{sec-conclusions}

The production of $D^{+}$ mesons has been measured in DIS at HERA in the kinematic region $5 < Q^{2} < 1000\gev^{2}, \, 0.02 < y < 0.7, \, 1.5 < p_{T}(D^{+}) < 15\gev$ and $|\eta(D^{+})| < 1.6$. The present results supersede the previous ZEUS $D^{+}$ measurement based on a subset of the data used in this analysis.  Predictions from NLO QCD describe the measured cross sections well. The charm contribution to the structure-function $F_2$ was extracted and agrees with that extracted from previous $D^{*}$ measurements. NLO QCD calculations describe the data well. 

The results presented here are of similar or higher precision than measurements previously published by ZEUS. The new precise data provide an improved check of pQCD and have the potential to constrain further the parton densities in the proton.

\section*{Acknowledgements}
\label{sec-ack}

\Zacknowledge

\vfill\eject

{
\def\bibname{\Large\bf References}
\def\refname{\Large\bf References}
\pagestyle{plain}
\ifzeusbst
  \bibliographystyle{./BiBTeX/bst/l4z_default}
\fi
\ifzdrftbst
  \bibliographystyle{./BiBTeX/bst/l4z_draft}
\fi
\ifzbstepj
  \bibliographystyle{./BiBTeX/bst/l4z_epj}
\fi
\ifzbstnp
  \bibliographystyle{./BiBTeX/bst/l4z_np}
\fi
\ifzbstpl
  \bibliographystyle{./BiBTeX/bst/l4z_pl}
\fi
{\raggedright
\bibliography{./BiBTeX/user/syn.bib,%
             ./BiBTeX/bib/l4z_articles.bib,%
             ./BiBTeX/bib/l4z_books.bib,%
             ./BiBTeX/bib/l4z_conferences.bib,%
             ./BiBTeX/bib/l4z_h1.bib,%
             ./BiBTeX/bib/l4z_misc.bib,%
             ./BiBTeX/bib/l4z_old.bib,%
             ./BiBTeX/bib/l4z_preprints.bib,%
             ./BiBTeX/bib/l4z_replaced.bib,%
             ./BiBTeX/bib/l4z_temporary.bib,%
             ./BiBTeX/bib/l4z_zeus.bib}}

\providecommand{\etal}{et al.\xspace}
\providecommand{\coll}{Coll.\xspace}
\catcode`\@=11
\def\@bibitem#1{%
\ifmc@bstsupport
  \mc@iftail{#1}%
    {;\newline\ignorespaces}%
    {\ifmc@first\else.\fi\orig@bibitem{#1}}
  \mc@firstfalse
\else
  \mc@iftail{#1}%
    {\ignorespaces}%
    {\orig@bibitem{#1}}%
\fi}%
\catcode`\@=12
\begin{mcbibliography}{10}

\bibitem{epj:c12:35}
ZEUS \coll, J.~Breitweg \etal,
\newblock Eur.\ Phys.\ J.{} {\bf C~12},~35~(2000)\relax
\relax
\bibitem{pr:d69:012004}
ZEUS \coll, S.~Chekanov \etal,
\newblock Phys.\ Rev.{} {\bf D~69},~012004~(2004)\relax
\relax
\bibitem{epj:c63:2009:2:171-188}
ZEUS Coll., S. Chekanov et al.,
\newblock Eur.\ Phys.\ J.{} {\bf C 63},~171~(2009)\relax
\relax
\bibitem{epj:c65:65-79}
ZEUS Coll., S. Chekanov et al.,
\newblock Eur.\ Phys.\ J.{} {\bf C~65},~65~(2010)\relax
\relax
\bibitem{jhep:2010:11:009}
ZEUS Coll., H. Abramowicz et al.,
\newblock JHEP{} {\bf 11},~1~(2010)\relax
\relax
\bibitem{pl:b528:199}
H1 \coll, C.~Adloff \etal,
\newblock Phys.\ Lett.{} {\bf B~528},~199~(2002)\relax
\relax
\bibitem{epj:c51:271}
H1 Coll., A. Aktas et al.,
\newblock Eur.\ Phys.\ J.{} {\bf C~51},~271~(2007)\relax
\relax
\bibitem{pl:b686:91}
H1 Coll., F.D. Aaron et al.,
\newblock Phys.\ Lett.{} {\bf B~686},~91~(2010)\relax
\relax
\bibitem{epj:c71:1769}
H1 Coll., F.D. Aaron et al.,
\newblock Eur.\ Phys.\ J.{} {\bf C~71},~1769~(2011)\relax
\relax
\bibitem{epj:c45:23}
H1 Coll., A. Aktas et al.,
\newblock Eur.\ Phys.\ J.{} {\bf C~45},~23~(2006)\relax
\relax
\bibitem{epj:c65:89}
H1 Coll., F.D. Aaron et al.,
\newblock Eur.\ Phys.\ J.{} {\bf C65},~89~(2010)\relax
\relax
\bibitem{epj:c71:1509}
H1 Coll., F.D. Aaron et al.,
\newblock Eur.\ Phys.\ J.{} {\bf C~71},~1509~(2011)\relax
\relax
\bibitem{tech:desy12-XXX:ccombination}
H1 and ZEUS Coll., H. Abramowicz et al.,
\newblock
\newblock Submitted to Eur. Phys. J. C. Available at
  arXiv:hep-ex/1211.1182\relax
\relax
\bibitem{pl:b699:345}
S. Alekhin and S. Moch,
\newblock Phys.\ Lett.{} {\bf B~699},~345~(2011)\relax
\relax
\bibitem{zeus:1993:bluebook}
ZEUS \coll, U.~Holm~(ed.),
\newblock {\em The {ZEUS} Detector}.
\newblock Status Report (unpublished), DESY (1993),
\newblock available on
  \texttt{http://www-zeus.desy.de/bluebook/bluebook.html}\relax
\relax
\bibitem{nim:a279:290}
N.~Harnew \etal,
\newblock Nucl.\ Inst.\ Meth.{} {\bf A~279},~290~(1989)\relax
\relax
\bibitem{npps:b32:181}
B.~Foster \etal,
\newblock Nucl.\ Phys.\ Proc.\ Suppl.{} {\bf B~32},~181~(1993)\relax
\relax
\bibitem{nim:a338:254}
B.~Foster \etal,
\newblock Nucl.\ Inst.\ Meth.{} {\bf A~338},~254~(1994)\relax
\relax
\bibitem{nim:a581:656}
A. Polini et al.,
\newblock Nucl.\ Inst.\ Meth.{} {\bf A~581},~656~(2007)\relax
\relax
\bibitem{nim:a309:77}
M.~Derrick \etal,
\newblock Nucl.\ Inst.\ Meth.{} {\bf A~309},~77~(1991)\relax
\relax
\bibitem{nim:a309:101}
A.~Andresen \etal,
\newblock Nucl.\ Inst.\ Meth.{} {\bf A~309},~101~(1991)\relax
\relax
\bibitem{nim:a321:356}
A.~Caldwell \etal,
\newblock Nucl.\ Inst.\ Meth.{} {\bf A~321},~356~(1992)\relax
\relax
\bibitem{nim:a336:23}
A.~Bernstein \etal,
\newblock Nucl.\ Inst.\ Meth.{} {\bf A~336},~23~(1993)\relax
\relax
\bibitem{desy-92-066}
J.~Andruszk\'ow \etal,
\newblock Preprint \mbox{DESY-92-066}, DESY, 1992\relax
\relax
\bibitem{zfp:c63:391}
ZEUS \coll, M.~Derrick \etal,
\newblock Z.\ Phys.{} {\bf C~63},~391~(1994)\relax
\relax
\bibitem{acpp:b32:2025}
J.~Andruszk\'ow \etal,
\newblock Acta Phys.\ Pol.{} {\bf B~32},~2025~(2001)\relax
\relax
\bibitem{nim:a565:572}
M.~Helbich \etal,
\newblock Nucl.\ Inst.\ Meth.{} {\bf A~565},~572~(2006)\relax
\relax
\bibitem{np:b374:36}
J.~Smith and W.L.~van Neerven,
\newblock Nucl.\ Phys.{} {\bf B~374},~36~(1992)\relax
\relax
\bibitem{pr:d57:2806}
B.W.~Harris and J.~Smith,
\newblock Phys.\ Rev.{} {\bf D~57},~2806~(1998)\relax
\relax
\bibitem{pr:d67:012007}
ZEUS \coll, S.~Chekanov \etal,
\newblock Phys.\ Rev.{} {\bf D~67},~012007~(2003)\relax
\relax
\bibitem{pl:b78:615}
V.G. Kartvelishvili, A.K. Likhoded and V.A. Petrov,
\newblock Phys.\ Lett.{} {\bf B~78},~615~(1983)\relax
\relax
\bibitem{thesis:lisovyi:2011}
M. Lisovyi,
\newblock Ph.D. Thesis, Universit\"at Hamburg, Report
  \mbox{DESY-THESIS-2011-033}, 2011\relax
\relax
\bibitem{jhep:04:082}
ZEUS Coll., S. Chekanov et al.,
\newblock JHEP{} {\bf 04},~082~(2009)\relax
\relax
\bibitem{epj:c59:589}
H1 Coll., F.D. Aaron et al.,
\newblock Eur.\ Phys.\ J.{} {\bf C~59},~589~(2009)\relax
\relax
\bibitem{jhep:0604:006}
M. Cacciari, P. Nason and C. Oleari,
\newblock JHEP{} {\bf 0604},~006~(2006)\relax
\relax
\bibitem{pr:d73:032002}
Belle Coll., R. Seuster et al.,
\newblock Phys.\ Rev.{} {\bf D~73},~032002~(2006)\relax
\relax
\bibitem{pr:d70:112001}
CLEO Coll., M. Artuso et al.,
\newblock Phys.\ Rev.{} {\bf D~70},~112001~(2004)\relax
\relax
\bibitem{upub:lohrmann:cfrac_arxiv}
E. Lohrmann,
\newblock {\em A summary of charm hadron production fractions} (unpublished).
\newblock Available at arXiv:hep-ex/1112.3757, 2011\relax
\relax
\bibitem{pr:d86:010001}
J. Beringer et al., Particle Data Group,
\newblock Phys.\ Rev.{} {\bf D~86},~010001~(2012)\relax
\relax
\bibitem{pr:d57:6871}
R.G.~Roberts and R.S.~Thorne,
\newblock Phys.\ Rev.{} {\bf D~57},~6871~(1998)\relax
\relax
\bibitem{np:b278:934}
G.C.~Collins and W.-K.~Tung,
\newblock Nucl.\ Phys.{} {\bf B~278},~934~(1986)\relax
\relax
\bibitem{cpc:86:147}
H.~Jung,
\newblock Comp.\ Phys.\ Comm.{} {\bf 86},~147~(1995)\relax
\relax
\bibitem{cpc:69:155}
A.~Kwiatkowski, H.~Spiesberger and H.-J.~M\"ohring,
\newblock Comp.\ Phys.\ Comm.{} {\bf 69},~155~(1992).
\newblock Also in {\it Proc.\ Workshop Physics at HERA}, eds. W.~Buchm\"{u}ller
  and G.Ingelman, (DESY, Hamburg, 1991)\relax
\relax
\bibitem{epj:c12:375}
CTEQ \coll, H.L.~Lai \etal,
\newblock Eur.\ Phys.\ J.{} {\bf C~12},~375~(2000)\relax
\relax
\bibitem{zfp:c11:169}
M.G.~Bowler,
\newblock Z.\ Phys.{} {\bf C~11},~169~(1981)\relax
\relax
\bibitem{tech:cern-dd-ee-84-1}
R.~Brun et al.,
\newblock {\em {\sc geant3}},
\newblock  CERN-DD/EE/84-1, CERN, 1987\relax
\relax
\bibitem{nim:a580:1257}
P.D.~Allfrey \etal,
\newblock Nucl.\ Inst.\ Meth.{} {\bf A~580},~1257~(2007)\relax
\relax
\bibitem{uproc:chep:1992:222}
W.H.~Smith, K.~Tokushuku and L.W.~Wiggers,
\newblock {\em Proc.\ Computing in High-Energy Physics (CHEP), Annecy, France,
  Sept. 1992}, C.~Verkerk and W.~Wojcik~(eds.), p.~222.
\newblock CERN, Geneva, Switzerland (1992).
\newblock Also in preprint \mbox{DESY 92-150B}\relax
\relax
\bibitem{proc:hera:1991:23}
S.~Bentvelsen, J.~Engelen and P.~Kooijman,
\newblock {\em Proc.\ Workshop on Physics at {HERA}}, W.~Buchm\"uller and
  G.~Ingelman~(eds.), Vol.~1, p.~23.
\newblock Hamburg, Germany, DESY (1992)\relax
\relax
\bibitem{thesis:briskin:1998}
G.M.~Briskin,
\newblock Ph.D.\ Thesis, Tel Aviv University, Report \mbox{DESY-THESIS
  1998-036}, 1998\relax
\relax
\bibitem{epj:c1:81}
ZEUS \coll, J.~Breitweg \etal,
\newblock Eur.\ Phys.\ J.{} {\bf C~1},~81~(1998)\relax
\relax
\bibitem{proc:epfacility:1979:391}
F.~Jacquet and A.~Blondel,
\newblock {\em Proceedings of the Study for an $ep$ Facility for {Europe}},
  U.~Amaldi~(ed.), p.~391.
\newblock Hamburg, Germany (1979).
\newblock Also in preprint \mbox{DESY 79/48}\relax
\relax
\bibitem{private:libov}
V. Libov, A. Spiridonov, private communication\relax
\relax
\bibitem{thesis:schoenberg:2010}
V. Sch\"onberg,
\newblock Dissertation, Universit\"at Bonn, Report \mbox{Bonn-IR-2010-05},
  2010,
\newblock available on
  \texttt{http://hss.ulb.uni-bonn.de/2010/2127/2127.pdf}\relax
\relax
\bibitem{epj:c69:347}
ZEUS Coll., H. Abramowicz et al.,
\newblock Eur.\ Phys.\ J.{} {\bf C~69},~347~(2010)\relax
\relax
\bibitem{epj:c71:1573}
ZEUS Coll., H. Abramowicz et al.,
\newblock Eur.\ Phys.\ J.{} {\bf C~71},~1573~(2011)\relax
\relax
\bibitem{misc:ichep10:168}
H1 and ZEUS Collaborations,
\newblock {\em HERAPDF1.5}.
\newblock Proceedings of the XXXV International Conference of High Energy
  Physics, Paris, 22--28 July 2010. PoS(ICHEP 2010)168, LHAPDF grid.
\newblock LHAPDF grid available on
  \texttt{https://www.desy.de/h1zeus/combined\_results/index.php?do=proton\_st%
ructure}\relax
\relax
\end{mcbibliography}
}
\vfill\eject

\begin{table}[p]
	\begin{center}
	\begin{tabular}{|c @{:} c|c c c|c c|}
	\hline
	\multicolumn{2}{|c|}{$Q^{2}$} & $d \sigma / d Q^{2}$ & $\Delta_{\mathrm{stat}}$ & $\Delta_{\mathrm{syst}}$ & $\mathcal{C}_{\rmrad}$ & $d \sigma_b / d Q^{2}$\\	
	\multicolumn{2}{|c|}{$(\Gev^{2})$} & \multicolumn{3}{c|}{$(\nb / \Gev^2)$} & & $(\nb / \Gev^2)$ \\
	\hline
         $5$ & $10$ & 0.382 & $\pm0.022$ & ${}^{+0.027}_{-0.017}$ & 1.018 & 0.007 \\
         $10$ & $20$ & 0.150 & $\pm0.007$ & ${}^{+0.008}_{-0.010}$ & 1.016 & 0.003 \\
         $20$ & $40$ & 0.047 & $\pm0.003$ & ${}^{+0.003}_{-0.004}$ & 1.020 & 0.002 \\
         $40$ & $80$ & 0.0108 & $\pm0.0008$ & ${}^{+0.0008}_{-0.0009}$ & 1.025 & 0.0006 \\
         $80$ & $200$ & 0.00192 & $\pm0.00020$ & ${}^{+0.00014}_{-0.00016}$ & 1.042 & 0.00016 \\
         $200$ & $1000$ & 0.000088 & $\pm0.000021$ & ${}^{+0.000006}_{-0.000007}$ & 1.113 & 0.000013 \\
        \hline	
	\hline
	\multicolumn{2}{|c|}{$y$} & $d \sigma / d y$ & $\Delta_{\mathrm{stat}}$ & $\Delta_{\mathrm{syst}}$& $\mathcal{C}_{\rmrad}$ & $d \sigma_b / d y$\\	
	\multicolumn{2}{|c|}{} & \multicolumn{3}{c|}{$(\nb)$} & & $(\nb)$ \\
	\hline
         0.02 & 0.1 & 16.9 & $\pm0.9$ & ${}^{+0.9}_{-0.8}$ & 1.038 & 0.1 \\
         0.1 & 0.2 & 13.4 & $\pm0.6$ & ${}^{+0.5}_{-0.5}$ & 1.022 & 0.3 \\
         0.2 & 0.3 & 8.5 & $\pm0.5$ & ${}^{+0.4}_{-0.4}$ & 1.025 & 0.3 \\
         0.3 & 0.4 & 6.2 & $\pm0.5$ & ${}^{+0.3}_{-0.3}$ & 1.016 & 0.3 \\
         0.4 & 0.5 & 4.0 & $\pm0.4$ & ${}^{+0.3}_{-0.2}$ & 1.008 & 0.2 \\
         0.5 & 0.7 & 2.2 & $\pm0.3$ & ${}^{+0.2}_{-0.2}$ & 0.999 & 0.2 \\
	\hline
	\end{tabular}
	\caption{Bin-averaged differential cross sections for $D^{+}$ production in the process $e p \rightarrow e' c \bar{c} X \rightarrow e' D^{+} X'$ in bins of $Q^{2}$ and $y$. The cross sections are given in the kinematic region \mbox{$5 < Q^{2} < 1000\gev^{2}$}, \mbox{$0.02 < y < 0.7$}, \mbox{$1.5 < p_{T}(D^{+}) < 15\gev$} and \mbox{$|\eta(D^{+})| < 1.6$}. The statistical and systematic uncertainties, $\Delta_{\mathrm{stat}}$ and $\Delta_{\mathrm{syst}}$, are presented separately. Normalisation uncertainties of $1.9\%$ and $2.1\%$ due to the luminosity and the branching-ratio measurements, respectively, were not included in $\Delta_{\mathrm{syst}}$. The correction factors to the QED Born level, $\mathcal{C}_{\rmrad}$ are also listed. For reference, the beauty cross section predicted by RAPGAP and scaled as described in the text, $\sigma_b$, are also shown.}
        \label{tab-xs_sd2}
	\end{center}
\end{table}
\begin{table}[p]
	\begin{center}
	\begin{tabular}{|c @{:} c|c c c|c c|}
	\hline
	\multicolumn{2}{|c|}{$p_{T}(D^{+})$} & $d \sigma / d p_{T}(D^{+})$ & $\Delta_{\mathrm{stat}}$ & $\Delta_{\mathrm{syst}}$& $\mathcal{C}_{\rmrad}$ & $d \sigma_b / d p_{T}(D^{+})$\\	
	\multicolumn{2}{|c|}{$(\Gev)$} & \multicolumn{3}{c|}{$(\nb / \Gev)$} & &$(\nb / \Gev)$\\
	\hline
         1.5 & 2.4 & 2.40 & $\pm0.26$ & ${}^{+0.14}_{-0.12}$ & 1.016 & 0.07\\
         2.4 & 3 & 1.44 & $\pm0.12$ & ${}^{+0.07}_{-0.05}$ & 1.020 & 0.05 \\
         3 & 4 & 1.00 & $\pm0.05$ & ${}^{+0.04}_{-0.04}$ & 1.023 & 0.03 \\
         4 & 6 & 0.396 & $\pm0.017$ & ${}^{+0.014}_{-0.013}$ & 1.029 & 0.011 \\
         6 & 15 & 0.0349 & $\pm0.0018$ & ${}^{+0.0011}_{-0.0010}$ & 1.054 & 0.0011\\
	\hline	
	\hline
	\multicolumn{2}{|c|}{$\eta(D^{+})$} & $d \sigma / d \eta(D^{+})$ & $\Delta_{\mathrm{stat}}$ & $\Delta_{\mathrm{syst}}$& $\mathcal{C}_{\rmrad}$ & $d \sigma_b / d \eta(D^{+})$\\	
	\multicolumn{2}{|c|}{} & \multicolumn{3}{c|}{$(\nb)$} & & $(\nb)$\\
	\hline
         $-1.6$ & $-0.8$ & 1.04 & $\pm0.09$ & ${}^{+0.06}_{-0.06}$ & 1.034 & 0.02 \\
         $-0.8$ & $-0.4$ & 1.67 & $\pm0.10$ & ${}^{+0.06}_{-0.06}$ & 1.025 & 0.05 \\
         $-0.4$ & $0.0$ & 1.70 & $\pm0.10$ & ${}^{+0.07}_{-0.05}$ & 1.023 & 0.05 \\
         $0.0$ & $0.4$ & 1.63 & $\pm0.10$ & ${}^{+0.07}_{-0.07}$ & 1.017 & 0.06 \\
         $0.4$ & $0.8$ & 1.84 & $\pm0.12$ & ${}^{+0.07}_{-0.08}$ & 1.013 & 0.06 \\
         $0.8$ & $1.6$ & 1.81 & $\pm0.16$ & ${}^{+0.09}_{-0.09}$ & 1.016 & 0.05 \\
	\hline
	\end{tabular}
	\caption{Bin-averaged differential cross sections for $D^{+}$ production in the process $e p \rightarrow e' c \bar{c} X \rightarrow e' D^{+} X'$ in bins of $p_{T}(D^{+})$ and $\eta(D^{+})$. Other details are as in \protect \tab{xs_sd2}.}
        \label{tab-xs_sd1}
	\end{center}
\end{table}

\begin{table}[bt]
	\begin{center}
	\renewcommand{\arraystretch}{1.2}
	\begin{tabular}{|c|c |c @{:} c|c c c|c c|}
	\hline
	Bin &$Q^{2}$ & \multicolumn{2}{c|}{$y$} & $d \sigma / d y$ & $\Delta_{\mathrm{stat}}$ & $\Delta_{\mathrm{syst}}$ & $\mathcal{C}_{\rmrad}$ & $d \sigma_b / d y$\\	
	& $(\Gev^{2})$ & \multicolumn{2}{c|}{} & \multicolumn{3}{c|}{$(\nb)$}& & $(\nb)$\\
	\hline	
         1 & \multirow{3}{*}{5 : 9} & 0.02 & 0.12 & 5.46 & $\pm0.59$ & ${}^{+0.46}_{-0.30}$ &1.026 & 0.04\\
         2 & & 0.12 & 0.32 & 3.40 & $\pm0.31$ & ${}^{+0.29}_{-0.16}$& 1.022 & 0.06 \\
         3 & & 0.32 & 0.7 & 1.18 & $\pm0.17$ & ${}^{+0.10}_{-0.08}$& 1.006 & 0.04 \\
	\hline
         4 & \multirow{3}{*}{9 : 23} & 0.02 & 0.12 & 7.02 & $\pm0.45$ & ${}^{+0.46}_{-0.49}$ &1.028 & 0.05\\
         5 & & 0.12 & 0.32 & 3.72 & $\pm0.23$ & ${}^{+0.21}_{-0.26}$& 1.017 & 0.09 \\
         6 & & 0.32 & 0.7 & 1.36 & $\pm0.14$ & ${}^{+0.09}_{-0.10}$& 0.998 & 0.06 \\
	\hline
         7 & \multirow{3}{*}{23 : 45} & 0.02 & 0.12 & 2.84 & $\pm0.27$ & ${}^{+0.19}_{-0.22}$ &1.040 & 0.03\\
         8 & & 0.12 & 0.32 & 1.63 & $\pm0.12$ & ${}^{+0.10}_{-0.12}$& 1.020 & 0.05 \\
         9 & & 0.32 & 0.7 & 0.609 & $\pm0.097$ & ${}^{+0.047}_{-0.053}$& 1.009 & 0.035 \\
	\hline
         10 & \multirow{3}{*}{45 : 100} & 0.02 & 0.12 & 1.14 & $\pm0.18$ & ${}^{+0.09}_{-0.10}$ &1.046 & 0.03\\
         11 & & 0.12 & 0.32 & 0.867 & $\pm0.083$ & ${}^{+0.063}_{-0.074}$& 1.024 & 0.050 \\
         12 & & 0.32 & 0.7 & 0.313 & $\pm0.052$ & ${}^{+0.032}_{-0.037}$& 1.012 & 0.033 \\
	\hline	
         13 & \multirow{2}{*}{100 : 1000} & 0.02 & 0.275 & 0.560 & $\pm0.085$ & ${}^{+0.031}_{-0.038}$ &1.117 & 0.033\\
         14 & & 0.275 & 0.7 & 0.231 & $\pm0.039$ & ${}^{+0.020}_{-0.022}$& 1.030 & 0.035 \\
	\hline
	\end{tabular}
	\caption{Bin-averaged differential cross sections for $D^{+}$ production in the process $e p \rightarrow e' c \bar{c} X \rightarrow e' D^{+} X'$ as a function of $y$ in five regions of $Q^{2}$. Other details are as in \protect \tab{xs_sd2}.}
        \label{tab-xs_dd}
	\end{center}
\end{table}

\begin{table}[h!]
        \begin{center}
        \renewcommand{\arraystretch}{1.3}
        \begin{tabular}{|c|*{11}{c}|}
        \hline
        Bin & $\delta_1$ & $\delta_2$ & $\delta_3$ & $\delta_4$ & $\delta_5$ & $\delta_6$ & $\delta_7$ & $\delta_8$ & $\delta_9$ & $\delta_{10}$ & $\delta_{11}$ \\
        \hline
         1 & ${}^{+6.4 \%}_{-0.0 \%}$ & ${}^{0.0 \%}_{0.0 \%}$ & ${}^{-0.9 \%}_{+0.9 \%}$ & ${}^{-0.3 \%}_{+0.3 \%}$ & ${}^{+1.3 \%}_{-0.0 \%}$ & ${}^{+0.5 \%}_{-1.5 \%}$ & ${}^{-0.3 \%}_{+0.3 \%}$ & ${}^{+0.9 \%}_{-0.9 \%}$ & ${}^{-3.6 \%}_{+3.6 \%}$ & ${}^{+2.4 \%}_{-3.1 \%}$ & ${}^{+1.7 \%}_{-1.7 \%}$ \\
         2 & ${}^{+7.6 \%}_{-1.8 \%}$ & ${}^{-0.1 \%}_{+0.0 \%}$ & ${}^{+0.3 \%}_{-0.0 \%}$ & ${}^{-1.3 \%}_{+1.3 \%}$ & ${}^{+0.1 \%}_{-0.0 \%}$ & ${}^{+0.9 \%}_{-2.4 \%}$ & ${}^{-0.2 \%}_{+0.2 \%}$ & ${}^{+1.4 \%}_{-1.4 \%}$ & ${}^{+0.9 \%}_{-0.9 \%}$ & ${}^{+1.8 \%}_{-2.3 \%}$ & ${}^{+1.6 \%}_{-1.6 \%}$ \\
         3 & ${}^{+4.9 \%}_{-0.0 \%}$ & ${}^{+0.2 \%}_{-0.0 \%}$ & ${}^{+0.6 \%}_{-0.5 \%}$ & ${}^{-1.6 \%}_{+1.6 \%}$ & ${}^{+0.5 \%}_{-0.0 \%}$ & ${}^{+1.0 \%}_{-2.4 \%}$ & ${}^{-0.6 \%}_{+0.6 \%}$ & ${}^{+2.2 \%}_{-2.2 \%}$ & ${}^{+5.0 \%}_{-5.0 \%}$ & ${}^{+2.0 \%}_{-2.6 \%}$ & ${}^{+1.6 \%}_{-1.6 \%}$ \\
         4 & ${}^{+1.5 \%}_{-0.0 \%}$ & ${}^{0.0 \%}_{0.0 \%}$ & ${}^{-0.1 \%}_{+0.3 \%}$ & ${}^{-0.2 \%}_{+0.2 \%}$ & ${}^{+2.1 \%}_{-0.0 \%}$ & ${}^{+0.4 \%}_{-1.0 \%}$ & ${}^{-0.1 \%}_{+0.1 \%}$ & ${}^{+0.7 \%}_{-0.7 \%}$ & ${}^{-2.3 \%}_{+2.3 \%}$ & ${}^{+4.8 \%}_{-6.1 \%}$ & ${}^{+1.6 \%}_{-1.6 \%}$ \\
         5 & ${}^{-0.6 \%}_{+0.0 \%}$ & ${}^{0.0 \%}_{0.0 \%}$ & ${}^{-0.1 \%}_{+0.3 \%}$ & ${}^{-1.1 \%}_{+1.1 \%}$ & ${}^{+0.3 \%}_{-0.0 \%}$ & ${}^{+0.8 \%}_{-2.2 \%}$ & ${}^{-0.2 \%}_{+0.2 \%}$ & ${}^{+1.1 \%}_{-1.1 \%}$ & ${}^{+0.4 \%}_{-0.4 \%}$ & ${}^{+4.7 \%}_{-6.0 \%}$ & ${}^{+1.5 \%}_{-1.5 \%}$ \\
         6 & ${}^{+0.3 \%}_{-0.0 \%}$ & ${}^{-1.6 \%}_{+2.2 \%}$ & ${}^{-0.2 \%}_{+0.1 \%}$ & ${}^{-1.2 \%}_{+1.2 \%}$ & ${}^{+0.2 \%}_{-0.0 \%}$ & ${}^{+0.7 \%}_{-1.6 \%}$ & ${}^{-1.3 \%}_{+1.3 \%}$ & ${}^{+2.3 \%}_{-2.3 \%}$ & ${}^{+3.7 \%}_{-3.7 \%}$ & ${}^{+3.4 \%}_{-4.3 \%}$ & ${}^{+1.6 \%}_{-1.6 \%}$ \\
         7 & ${}^{+0.0 \%}_{-0.2 \%}$ & ${}^{0.0 \%}_{0.0 \%}$ & ${}^{-0.4 \%}_{+0.3 \%}$ & ${}^{-0.1 \%}_{+0.1 \%}$ & ${}^{+0.6 \%}_{-0.0 \%}$ & ${}^{+0.5 \%}_{-1.4 \%}$ & ${}^{+0.1 \%}_{-0.1 \%}$ & ${}^{+0.8 \%}_{-0.8 \%}$ & ${}^{-2.3 \%}_{+2.3 \%}$ & ${}^{+5.5 \%}_{-7.1 \%}$ & ${}^{+1.5 \%}_{-1.5 \%}$ \\
         8 & ${}^{0.0 \%}_{0.0 \%}$ & ${}^{0.0 \%}_{0.0 \%}$ & ${}^{-0.2 \%}_{+0.4 \%}$ & ${}^{-0.3 \%}_{+0.3 \%}$ & ${}^{+0.6 \%}_{-0.0 \%}$ & ${}^{+0.1 \%}_{-0.2 \%}$ & ${}^{-0.6 \%}_{+0.6 \%}$ & ${}^{+2.7 \%}_{-2.7 \%}$ & ${}^{0.0 \%}_{0.0 \%}$ & ${}^{+5.0 \%}_{-6.5 \%}$ & ${}^{+1.3 \%}_{-1.3 \%}$ \\
         9 & ${}^{+0.0 \%}_{-0.6 \%}$ & ${}^{-1.5 \%}_{+1.3 \%}$ & ${}^{+1.4 \%}_{-0.0 \%}$ & ${}^{-0.1 \%}_{+0.1 \%}$ & ${}^{-0.6 \%}_{+0.0 \%}$ & ${}^{+0.8 \%}_{-2.1 \%}$ & ${}^{-0.2 \%}_{+0.2 \%}$ & ${}^{+3.7 \%}_{-3.7 \%}$ & ${}^{+2.8 \%}_{-2.8 \%}$ & ${}^{+5.2 \%}_{-6.7 \%}$ & ${}^{+1.5 \%}_{-1.5 \%}$ \\
         10 & ${}^{0.0 \%}_{0.0 \%}$ & ${}^{0.0 \%}_{0.0 \%}$ & ${}^{-0.9 \%}_{+0.4 \%}$ & ${}^{0.0 \%}_{0.0 \%}$ & ${}^{+2.7 \%}_{-0.0 \%}$ & ${}^{+0.8 \%}_{-2.1 \%}$ & ${}^{-0.1 \%}_{+0.1 \%}$ & ${}^{+0.9 \%}_{-0.9 \%}$ & ${}^{-1.8 \%}_{+1.8 \%}$ & ${}^{+6.1 \%}_{-7.8 \%}$ & ${}^{+1.3 \%}_{-1.3 \%}$ \\
         11 & ${}^{-0.0 \%}_{+0.1 \%}$ & ${}^{0.0 \%}_{0.0 \%}$ & ${}^{-0.6 \%}_{+0.4 \%}$ & ${}^{0.0 \%}_{0.0 \%}$ & ${}^{+1.1 \%}_{-0.0 \%}$ & ${}^{+0.2 \%}_{-0.6 \%}$ & ${}^{-0.1 \%}_{+0.1 \%}$ & ${}^{+2.6 \%}_{-2.6 \%}$ & ${}^{-0.3 \%}_{+0.3 \%}$ & ${}^{+6.1 \%}_{-7.9 \%}$ & ${}^{+1.1 \%}_{-1.1 \%}$ \\
         12 & ${}^{+0.0 \%}_{-0.3 \%}$ & ${}^{-1.4 \%}_{+0.2 \%}$ & ${}^{+1.0 \%}_{-0.0 \%}$ & ${}^{0.0 \%}_{0.0 \%}$ & ${}^{+0.4 \%}_{-0.0 \%}$ & ${}^{+0.5 \%}_{-1.2 \%}$ & ${}^{-0.6 \%}_{+0.6 \%}$ & ${}^{+5.2 \%}_{-5.2 \%}$ & ${}^{+2.3 \%}_{-2.3 \%}$ & ${}^{+7.9 \%}_{-10.1 \%}$ & ${}^{+1.3 \%}_{-1.3 \%}$ \\
         13 & ${}^{+0.0 \%}_{-0.5 \%}$ & ${}^{0.0 \%}_{0.0 \%}$ & ${}^{-0.3 \%}_{+0.7 \%}$ & ${}^{0.0 \%}_{0.0 \%}$ & ${}^{-0.1 \%}_{+0.0 \%}$ & ${}^{+1.3 \%}_{-3.2 \%}$ & ${}^{0.0 \%}_{0.0 \%}$ & ${}^{+2.6 \%}_{-2.6 \%}$ & ${}^{-1.5 \%}_{+1.5 \%}$ & ${}^{+3.8 \%}_{-4.9 \%}$ & ${}^{+1.0 \%}_{-1.0 \%}$ \\
         14 & ${}^{-0.0 \%}_{+0.1 \%}$ & ${}^{-1.2 \%}_{+0.0 \%}$ & ${}^{-1.3 \%}_{+0.9 \%}$ & ${}^{-0.5 \%}_{+0.5 \%}$ & ${}^{+0.2 \%}_{-0.0 \%}$ & ${}^{+0.5 \%}_{-1.3 \%}$ & ${}^{-0.2 \%}_{+0.2 \%}$ & ${}^{+5.5 \%}_{-5.5 \%}$ & ${}^{+0.5 \%}_{-0.5 \%}$ & ${}^{+5.8 \%}_{-7.5 \%}$ & ${}^{+1.0 \%}_{-1.0 \%}$ \\
        \hline
        \end{tabular}
        \caption{Contributions of individual sources of systematics for the differential cross sections in bins of $y$ in five ranges of $Q^2$. The first column gives the bin number that is consistent with \protect \tab{xs_dd}. The systematic variation numbering is consistent with \Sect{syst}. Normalisation uncertainties $\delta_{12} \rnge \delta_{15}$ are not shown.}
        \label{tab-syst_dd}
        \end{center}
\end{table}

\begin{table}[bt]
	\begin{center}
	\renewcommand{\arraystretch}{1.15}
	\begin{tabular}{|c|c|c c c c|}
	\hline
	$Q^{2}$ & $x$ & $F_2^{c\bar{c}}$ & $\Delta_{\mathrm{stat}}$ & $\Delta_{\mathrm{syst}}$ & $\Delta_{\mathrm{theo}}$ \\	
	$(\Gev^{2})$ & & & & &\\
	\hline
         \multirow{3}{*}{6.5} &  0.00016 & 0.238 & $\pm0.033$ & ${}^{+0.020}_{-0.017}$ & ${}^{+0.033}_{-0.041}$ \\
          &      0.00046 & 0.147 & $\pm0.013$ & ${}^{+0.013}_{-0.007}$ & ${}^{+0.023}_{-0.013}$ \\
          &      0.00202 & 0.073 & $\pm0.008$ & ${}^{+0.006}_{-0.004}$ & ${}^{+0.010}_{-0.009}$ \\
        \hline
         \multirow{3}{*}{20.4} &         0.0005 & 0.363 & $\pm0.037$ & ${}^{+0.025}_{-0.025}$ & ${}^{+0.032}_{-0.049}$ \\
          &      0.00135 & 0.209 & $\pm0.013$ & ${}^{+0.012}_{-0.014}$ & ${}^{+0.015}_{-0.013}$ \\
          &      0.0025 & 0.170 & $\pm0.011$ & ${}^{+0.011}_{-0.012}$ & ${}^{+0.017}_{-0.012}$ \\
        \hline
         \multirow{3}{*}{35} &   0.0008 & 0.377 & $\pm0.060$ & ${}^{+0.029}_{-0.033}$ & ${}^{+0.023}_{-0.027}$ \\
          &      0.0014 & 0.275 & $\pm0.021$ & ${}^{+0.017}_{-0.020}$ & ${}^{+0.016}_{-0.014}$ \\
          &      0.0034 & 0.211 & $\pm0.020$ & ${}^{+0.014}_{-0.017}$ & ${}^{+0.013}_{-0.020}$ \\
        \hline
         \multirow{3}{*}{60} &   0.0015 & 0.265 & $\pm0.044$ & ${}^{+0.027}_{-0.032}$ & ${}^{+0.015}_{-0.013}$ \\
          &      0.0032 & 0.212 & $\pm0.020$ & ${}^{+0.015}_{-0.018}$ & ${}^{+0.010}_{-0.010}$ \\
          &      0.008  & 0.138 & $\pm0.022$ & ${}^{+0.010}_{-0.012}$ & ${}^{+0.013}_{-0.009}$ \\
        \hline
         \multirow{2}{*}{200} &  0.005 & 0.215 & $\pm0.036$ & ${}^{+0.018}_{-0.021}$ & ${}^{+0.013}_{-0.008}$ \\
          &      0.013 & 0.175 & $\pm0.026$ & ${}^{+0.010}_{-0.012}$ & ${}^{+0.011}_{-0.008}$ \\
        \hline
	\end{tabular}
	\caption{The values of \ftwocc at each $Q^2$ and $x$. The statistical ($\Delta_{\mathrm{stat}}$), systematic ($\Delta_{\mathrm{syst}}$) and theoretical ($\Delta_{\mathrm{theo}}$) uncertainties are given separately. Further uncertainties of $1.9\%$ and $2.1\%$ due to the luminosity and the branching-ratio measurements, respectively, were not included in $\Delta_{\mathrm{syst}}$. The theoretical uncertainty, $\Delta_{\mathrm{theo}}$, represents the uncertainty due to the extrapolation.}
        \label{tab-f2c}
	\end{center}
\end{table}

\begin{table}[bt]
	\begin{center}
	\renewcommand{\arraystretch}{1.15}
	\begin{tabular}{|c|c|c c c c|}
	\hline
	$Q^{2}$ & $x$ & $\sigma^{c\bar{c}}_{\mathrm{red}}$ & $\Delta_{\mathrm{stat}}$ & $\Delta_{\mathrm{syst}}$ & $\Delta_{\mathrm{theo}}$ \\	
	$(\Gev^{2})$ & & \multicolumn{4}{c|}{$(\nb)$}\\
	\hline
         \multirow{3}{*}{7} &    0.00016 & 0.249 & $\pm0.035$ & ${}^{+0.021}_{-0.017}$ & ${}^{+0.034}_{-0.043}$ \\
          &      0.00046 & 0.155 & $\pm0.014$ & ${}^{+0.013}_{-0.007}$ & ${}^{+0.025}_{-0.013}$ \\
          &      0.00202 & 0.077 & $\pm0.008$ & ${}^{+0.007}_{-0.004}$ & ${}^{+0.011}_{-0.009}$ \\
        \hline
         \multirow{3}{*}{18} &   0.0005 & 0.336 & $\pm0.034$ & ${}^{+0.023}_{-0.024}$ & ${}^{+0.029}_{-0.045}$ \\
          &      0.00135 & 0.198 & $\pm0.012$ & ${}^{+0.011}_{-0.014}$ & ${}^{+0.014}_{-0.012}$ \\
          &      0.0025 & 0.161 & $\pm0.010$ & ${}^{+0.011}_{-0.011}$ & ${}^{+0.017}_{-0.012}$ \\
        \hline
         \multirow{3}{*}{32} &   0.0008 & 0.352 & $\pm0.056$ & ${}^{+0.027}_{-0.031}$ & ${}^{+0.022}_{-0.025}$ \\
          &      0.0014 & 0.263 & $\pm0.020$ & ${}^{+0.017}_{-0.019}$ & ${}^{+0.015}_{-0.013}$ \\
          &      0.0034 & 0.203 & $\pm0.020$ & ${}^{+0.013}_{-0.016}$ & ${}^{+0.013}_{-0.019}$ \\
        \hline
         \multirow{3}{*}{60} &   0.0015 & 0.259 & $\pm0.043$ & ${}^{+0.026}_{-0.031}$ & ${}^{+0.015}_{-0.013}$ \\
          &      0.0032 & 0.211 & $\pm0.020$ & ${}^{+0.015}_{-0.018}$ & ${}^{+0.010}_{-0.010}$ \\
          &      0.008 & 0.138 & $\pm0.022$ & ${}^{+0.010}_{-0.012}$ & ${}^{+0.013}_{-0.009}$ \\
        \hline
         \multirow{2}{*}{200} &  0.005 & 0.210 & $\pm0.035$ & ${}^{+0.018}_{-0.020}$ & ${}^{+0.013}_{-0.008}$ \\
          &      0.013 & 0.175 & $\pm0.026$ & ${}^{+0.010}_{-0.012}$ & ${}^{+0.011}_{-0.008}$ \\
        \hline
	\end{tabular}
	\caption{The values of reduced cross sections, $\sigma^{c\bar{c}}_{\mathrm{red}}$, as a function of $Q^2$ and $x$. Other details are as in \protect \tab{f2c}.}
        \label{tab-sigmac}
	\end{center}
\end{table}

\vfill\eject

%
%
\end{document}